\documentclass[12pt]{article}

\usepackage{graphicx}
\usepackage{natbib}

\def    \be            {\begin{equation}}
\def    \ee            {\end{equation}}
\def    \bea           {\begin{eqnarray}}
\def    \eea           {\end{eqnarray}}

\newcommand{\ignorar}[1]{}
 
\newcommand{\bi}{\begin{itemize}}
\newcommand{\ei}{\end{itemize}}

\begin{document}

\begin{center}
{\Large \bf Theory of input spike
\vspace{0.2cm}

auto- and cross-correlations and their effect 
\vspace{0.4cm}

on the response of spiking neurons}

\vspace{1cm}

{\bf \sc Rub\'en Moreno-Bote (1,2),  Alfonso Renart (2,3) \\
and N\'estor Parga (2)}

\vspace{1cm}
 
(1) Present Address: Center for Neural Science, New York University, \\ New York, NY
10003-6621, USA.

(2) Dept. de F\'{\i}sica Te\'orica. Universidad Aut\'onoma de Madrid, 
\\Cantoblanco 28049, Madrid, Spain

(3) Present Address: Center for Molecular and Behavioral Neuroscience,
\\Rutgers, The State University of New Jersey, 197 University Avenue,
\\Newark, NJ 07102, USA

\end{center}

\begin{abstract} 
Spike correlations between neurons are ubiquitous in the cortex, but their role is at present
not understood. Here we describe the firing response of a leaky integrate-and-fire neuron
(LIF) when it receives a temporarily correlated input generated by
presynaptic correlated neuronal populations. Input correlations are characterized
in terms of the firing rates, Fano factors, correlation coefficients and correlation
timescale of the neurons driving the target neuron. We show that the sum of
the presynaptic spike trains cannot be well described by a Poisson process. 
In fact, the total input current has a non trivial 
two-point correlation function described by two main
parameters: the correlation timescale (how precise the input correlations are in
time), and the correlation magnitude (how strong they are).
Therefore, the total current generated by the input spike trains 
is not well described by a white noise Gaussian process. 
Instead, we model the total current 
as a colored Gaussian process with the same mean and two-point 
correlation function, leading
to the formulation of the problem in terms of a Fokker-Planck equation. Solutions
of the output firing rate are found in the limit of short and long correlation time
scales. The solutions described here expand and improve our previous results 
\citep{Mor+02} by presenting new analytical expressions for the output firing
rate for general IF neurons, extending the validity of the results for arbitrarily
large correlation magnitude, and by describing the differential effect of correlations
on the mean driven or noise dominated firing regimes.
Also the details of this novel formalism are given
here for the first time. We employ numerical simulations
to confirm the analytical solutions and to study the firing response to sudden
changes in the input correlations. 
We expect this formalism to be useful for 
the study of correlations in neuronal networks and their role in neural processing
and information transmission.
\end{abstract}

\section{Introduction}

A major problem in neuroscience is to understand the way neurons
communicate with each other. Because neurons in the cortex are densely
connected and share common inputs
\citep{Whi89,Bra+91}, some degree of correlation between their
discharges is unavoidable. Indeed, correlations in the spiking activity of neurons
are routinely observed throughout the cortex
(\citep{Zoh+94,deC+96,Lee+98-var,Unrey+99-syn,Bair+01-tauc}; for a review
see \citet{Sal+01a,Averbeck+04-syn}).
Correlations could have an important functional role, as the temporal
synchronization of neuronal activity has been shown to correlate with
particular states of behaving animals
\citep{Vaadia+95,Rie+97,Fries+97,Ste+00,Fri+01}. From a more traditional
point of view, correlations have been considered as a coding dimension
independent of the firing rate \citep{deC+96,Weh+99,Lau+01}.
However it remains still controversial whether correlated activity
has a role in coding, or whether its main role is as a gating mechanism of the flow
of information in cortical circuits \citep{Sal+01a,Averbeck+04-syn}.

Before the functional role of correlations can be addressed, a prime
question to solve is how correlations affect the firing properties of
neurons. Previous work in this direction has revealed
that neurons can be very sensitive even to weak correlations in their
inputs \citep{Bur+99,Fen+00,Sal+00}. However, in most of these studies,
only zero time lag correlated inputs ({\em perfect synchronization})
has been used. This means that when one spike arrives at one presynaptic terminal, 
another spike is more likely to be found {\em at the same time}
in other presynaptic terminal. This perfect synchrony is not expected to be
exhibited by real neuronal systems, given their finite temporal
precision. Instead, synchrony with a non zero time precision $\tau_c$
seems to be the realistic case, with $\tau_c \sim 15ms$ in monkey
primary auditory cortex \citep{deC+96}, $\tau_c \sim 5ms$ in primary
visual cortex of strabismic cats \citep{Fries+97} (in this case the
cross-correlogram is accompanied by an oscillatory pattern), $\tau_c$
with very broad values ranging from less than $15ms$ to more than
$200ms$ mediating interactions between areas $V1$ and $V2$ in monkeys
\citep{Nowak+99-corr}, or $\tau_c \sim 10ms$ in the monkey visual area
MT \citep{Bair+01-tauc}. In this case, if a spike arrives at time
$t=0$ at a presynaptic terminal, another spike is more, or less, likely 
than the chance level determined by the firing rate, to arrive within a 
time $\tau_c$ around $t=0$ at other (or the same) terminal.

We have shown previously \citep{Mor+02} 
that the total current to a neuron
generated by exponentially correlated afferent spike trains can be
described (among other parameters) by the {\em correlation time},
$\tau_c$, and the {\em correlation magnitude}, $\alpha$ (see
definitions in Section (\ref{sec:current})). Each parameter carries
important information about the characteristics of the input correlations
(either temporal or intensity information). Intuitively, a short correlation time
$\tau_c$ means that afferent spikes synchronize within short time
windows of size $\tau_c$.
Decreasing $\tau_c$ will enhance the temporal
precision of correlations. The correlation
magnitude, $\alpha$, roughly represents how many spikes are expected
above chance in a time window $\tau_c$ given that there was a spike centered 
in that time window. Therefore, it is a measure of the intensity of the
correlations. For uncorrelated spike trains $\alpha=0$, while for
positively correlated spike trains $\alpha>0$, and for negatively correlated 
$\alpha<0$. As we will show, the correlation time and magnitude can also be related
to the autocorrelograms (ACGs) and cross-correlograms (CCGs) of recorded spike
trains. The correlation time measures the typical width of the CCG, while
the correlation magnitude is proportional to the area under the CCG curve.

Both $\tau_c$ and $\alpha$ can affect the neuron's firing response in complicated
ways. Separating their effects was crucial in our previous work \citep{Mor+02},
where the effects of changing the timescale and the magnitude of the input
correlations could be studied independently. In particular, one of the main
qualitative results was that, if $\alpha$ is kept constant,
neurons are sensitive to input correlations only when the correlation time is
shorter than the membrane time constant.
\footnote{ This mechanism is consistent with {\it
    coincidence detection} \citep{Abe+82,Ber+91,Sof+93,Sof94}.
   Note, however, that these authors consider input spike
  coincidence detection in the sub-millisecond range, while our results more
  generally concern the effect of correlation timescale of any size on a neuron
  with any membrane time constant}.

The main problem studied in this paper is schematized in Fig.(\ref{fig:scheme}) and  
can be summarized as follows: What is the effect of the magnitude and the 
timescale of the input spike correlations on the neuron firing response? 
We answer this question by addressing consecutively several subproblems. 
First, after presenting the model (Sec \ref{sec:model}), we describe the
statistical properties of the afferent spike trains which 
drive a LIF neuron (Sec \ref{sec:current}). The spike trains are characterized 
in terms of their firing rates, Fano factors, correlation coefficients and correlation
timescale, and are assumed to have exponential auto- and
cross-correlations.
Correlated and uncorrelated Poisson spike trains are 
just special cases of these. The total current generated by the sum of the spike trains is
described up to second order statistics (the two-point correlation 
function), and shows exponential correlations (Sec \ref{sec:cross-corr}).
Second, to solve the difficulties presented by the non-Markovian 
character of the input statistics, we seek to transform this input into a colored
Gaussian input with the same mean and two-point correlation function as
those generated by the original current.
Two different Markovian stochastic processes that generate this colored Gaussian
input are found (Sec \ref{sec:FPequations}). 
Then, we obtain the Fokker-Plank equations (FPEs) 
associated to each of these two processes and the voltage of the neuron  
(Secs \ref{subsec:pos_neg_corr} and \ref{subsec:pos_corr}).
Third, the output firing rate is obtained by solving the FPEs in the
limits of short and long values of the correlation timescale 
compared to the membrane time constant of the neuron (Sec \ref{sec:solution}).  
At this point we give a brief summary of the analytical expressions and their ranges of
validity (Sec \ref{subsec:summary} and Table \ref{table1}). 
An interpolation is then employed to join the two
limits, and the analytical results are compared with numerical simulations
(Sec \ref{subsec:results}). 
Finally, we also show that neurons can
track fast changes in input correlations
(Sec  \ref{subsec:transient}). In the discussion section (Sec \ref{sec:discussion}) 
we summarize the main results and discuss possible applications. 
Several computational details are provided in a set of appendices.

Some of these results have been previously published in a brief format
\citep{Mor+02}. In the current work we extend the analytical techniques,
obtain new results and present a more pedagogical version of our 
work to facilitate the use of the mathematical expressions as well as the
understanding of their derivation. 
In particular, a more general expression
for the output firing rate is found in the presence of exponentially correlated
input spike trains that is valid for long $\tau_c$ and for all positive $\alpha$
(Sec 5.2).
If the limit of small $\alpha$ is taken, this new expression 
becomes that found in \citep{Mor+02} in the case of long $\tau_c$, and
therefore generalizes and extends the latter for large correlation magnitudes.
The effect of input correlations in the mean driven and noise dominated
input regimes is found to be different, and those peculiarities are discussed here
(Sec \ref{subsec:results}).


\section{Model}
\label{sec:model}


We consider a LIF neuron with membrane potential $V(t)$ and membrane
time constant $\tau_m$. In the absence of input, the voltage decays
exponentially 
toward the resting potential (here $V=0$).
In the presence of synaptic current, $I(t)$, the membrane potential evolves
according to the equation

\begin{equation}
\dot{V}(t) = - \frac{V(t)}{\tau_m} + I(t) \; .
\label{IF_equation}
\end{equation}

\noindent
In the model, a spike is generated whenever the membrane potential $V(t)$
reaches a threshold value $\Theta$. Following the spike, the
potential is reset to a value $H$, from where, after an absolute
refractory period $\tau_{ref}$, the neuron can start integrating
the synaptic current again.

We work in the limit of infinitely fast synaptic
time constants, in which individual synaptic currents are represented
by delta functions.  Thus, the afferent current $I(t)$ is

\begin{equation}
I(t)=J_{E} \; \sum_{i=1}^{N_E} \sum_{k} \delta(t-t_{i}^{k}) - 
     J_{I} \; \sum_{j=1}^{N_I} \sum_{l} \delta(t-t_{j}^{l})  \; ,
\label{eq:current-deltas}
\end{equation}

\noindent 
where $t_{i(j)}^{k(l)}$ represents the arrival time of the $k$-th
($l$-th) spike from the $i$-th excitatory ($j$-th inhibitory)
presynaptic neuron, and $N_{E(I)}$ and $J_{E(I)}$ 
represent, respectively, the number of inputs and the size of the postsynaptic
potentials from the excitatory (inhibitory) afferent populations.

We are interested in the case of stationary input statistics, so that the
input firing rates do not depend on time (but see our simulation results
for the case on non-stationary statistics in Sec. \ref{subsec:transient}). 
Therefore, assuming that the excitatory and inhibitory presynaptic neurons fire at rates
$\nu_E$ and $\nu_I$ respectively, the mean current $\left< I(t) \right>$ is computed as

\be
  \mu = \left< I(t) \right> = N_E \; J_E \; \nu_E - N_I \; J_I \; \nu_I \;\; .
  \label{eq:mu}
\ee

\noindent
This result is independent of the statistics of the afferent spike trains. For
example, the mean current generated by correlated or independent Poisson spike trains
is exactly the same, provided that the processes are stationary and described
by the same firing rates. However, the second-order statistics of the current
will be very sensitive to the second order statistical properties of the individual spike
trains (e.g., their pair-wise correlations).
In the next section we determine
the two-point correlation function in terms of the
statistical properties of the presynaptic spike trains.


\section{Second-order statistical properties of the current}
\label{sec:current}

\subsection{Auto-correlograms}

This section is devoted to the description of the second order statistical
properties of each individual spike train impinging on the LIF neuron. In the next section,
we will consider the second order statistical properties of pairs of those spike trains.
Here, we first define the Fano factor of the spike count of each input train.
Then we introduce the auto-correlation function in the case of an 
exponentially correlated spike train. 
Finally, we show that the parameters defining the exponential 
auto-correlation function can be expressed in terms of the firing rate, 
Fano factor and correlation time of the spike train.

Most theoretical models have considered afferent spike trains (see eq.
(\ref{eq:current-deltas})) as stochastic Poisson processes
(see e.g. \citep{Ric77,Tuc88,Bru+98b,Fen+00,Nykamp+01,LaCamera+04,Richardson+05}). 
In this work, we relax this assumption.
The Fano Factor is often used to quantify the reliability 
of neuronal discharge.
The Fano factor of the spike count 
in a time window $T$ is defined as 
the ratio between the variance of the spike count and the 
mean number of spikes in that time window, that is, 

\be
 F_N(T)  =  \frac{ \sigma_N^2(T) }{\left< N(T) \right>}
           = \frac {\left< (N(T)-\left< N(T)\right>)^2 \right> }
                 {\left< N(T) \right>}
                 \; ,
 \label{eq:F_NTdef}
\ee

\noindent
where $N(T)$ is the number of spikes counted in the time window $T$
in each trial and brackets denote an average over trials. Note that, in practice, the
mean and variance can also be computed using a single long spike train (with stationary
firing rate) obtained in a single trial, where now the average is obtained
using non-overlapping consecutive time windows instead of several trials. 
In either case, typically the time window $T$ is taken to be large, so that
at least tens of spikes are observed on average.
A Poisson spike train has a Fano factor equal to one. However, 
Fano factors calculated from spike trains obtained from
electrophysiological recordings {\em in vivo} usually exceed one, 
laying in the interval $F_N \sim 1 - 1.5$
throughout the cerebral cortex
\citep{Dea81,Sof+93,Alb93,Sha+98,Compte+03-F_N}, which is inconsistent
with the Poisson hypothesis (see also \citep{Ama+06}).

Another important second-order statistical property of individual spike trains is
the joint probability density of having spikes belonging to that same spike train
at two times, $t$ and $t'$, denoted $P(t,t')$. In 
fact, from it one can derive any other second-order statistical
quantity, such as the Fano factor (see below).
For a Poisson spike train with rate $\nu$, 
$P(t,t')$ is a delta function at zero-time lag 
and flat otherwise, as

\be
  P_{Poisson}(t,t') = \nu \delta(t-t') + \nu^2 
  \; .
  \label{auto-Poisson}
\ee

\noindent
The delta function at $t=t'$ serves to define $P(t,t')$ at all times; trivially,
the probability density of having a spike at time $t$ and a spike at time $t'=t$
is just the delta multiplied by the spike rate in that train, i.e., $\nu \delta(t-t') $;
in other words, the presence
of one spike is informative of the presence of a 
spike at that time (the same spike). For
non-zero time lags ($t \neq t'$), this probability 
is just the product of the probability densities
of having spikes at two different times, that is, $\nu^2$. For a general spike train
we define the autocorrelation function as the quantity 

\be
  C(t,t')=P(t,t')-\nu^2 , 
\ee

\noindent
that is, the joint probability
density of having spikes at times $t$ and $t'$, from which the probability of finding
them by chance (i.e., the rate to the square) is subtracted.

While Poisson trains have an autocorrelation with a single delta function at
time lag zero and zero otherwise (i.e. $C_{Poisson}(t,t') = \nu \delta(t-t') $), 
auto-correlograms obtained from electrophysiological recordings
show a decaying peak at non-zero time lags (disregarding refractory
effects) sometimes together with a damped oscillatory pattern. 
A centered decaying peak in an auto-correlogram means that spikes tend to
occur close together in time, forming groups of several spikes.
Experimental auto-correlograms with a single peak and without
oscillations can be fitted to an exponential function (e.g.
\cite{Bair+01-tauc}). We therefore consider stochastic spike trains with
exponential autocorrelations with timescale $\tau_c$

\begin{eqnarray}
C_{p}(t,t')&\equiv& \left< \left( \sum_{k}
\delta(t-t_{i}^{k}) -\nu_p \right) \left( \sum_{k'}
\delta(t'-t_{i}^{k'}) -\nu_p \right) \right>  
\nonumber 
\\
 &=& \left< \sum_{k,k'}
\delta(t-t_{i}^{k})\delta(t'-t_{i}^{k'}) \right> -\nu_p^2 
\nonumber 
\\ 
&=&
\nu_p \delta(t-t') + \nu_p \left(\frac{F_p-1}{2 \tau_{c}}\right) \;
e^{-\frac{\mid t-t'\mid}{\tau_c}}  \; ,
\label{autocorrelation}
\end{eqnarray}

\noindent
as illustrated in Fig.(\ref{fig:FN} B). Since we assume that the input
statistics is stationary, the input firing rates are time independent
and the auto-correlation function only depends on time through the
quantity $|t-t'|$,
Here $p=E,I$; $\nu_p$ and $F_p$ are the firing rate and the Fano
factor of the spike count (for infinitely long time windows) of the
individual trains coming from population $p$ 
\footnote{For renewal spike trains, the
  Fano factors in the above equations are related to the coefficients
  of variation of their inter-spike-intervals, $CV_p$, as $F_p=CV_p^2$. Note
  nevertheless that our formalism does not require that afferent spike
  trains are renewal.}. 
The connected two-point
correlation function defined above is the joint probability density of
finding one spike at time $t$ and another at $t'$ within the same
spike train, from where the probability of observing
them by chance, $\nu^2_p$, is subtracted. Note that this function has two
contributions: a delta function at zero time lag, coming from the fact
that spikes are point events, and an exponential dependence measuring
the excess probability of finding a spike at $t'$ when it is known
that there is another spike at $t$. 
While normally spikes in the same train are positively correlated
($F_N > 1$), the auto-correlogram in eq. (\ref{autocorrelation}) also
describes uncorrelated ($F_N = 1$, Poisson) and negatively correlated
spikes ($F_N < 1$).  With the parameterization we have chosen, fixing
the Fano factor and changing the correlation time does not keep fixed
the amplitude of the exponential term in eq. (\ref{autocorrelation}).
However, this choice allows us to fix the variance of the spike count
in a long time window for each individual spike train while varying the
timescale of its correlations. To make this clearer, consider the
total number of presynaptic spikes arriving from the spike train $i$
of the population $p$ during a time window $T$, which is written as

\be
 N(T)= \int_{0}^{T} dt \sum_{k} \delta(t-t_{i}^{k})
 \; .
 \nonumber
\ee

\noindent
Notice that, since the arrival times $t_{i}^{k}$ are random in such a
way that the train has the autocorrelation of eq.
(\ref{autocorrelation}), the number $N(T)$ is a random variable. Its
mean value is 

\be
 \left< N_p(T) \right>= \left< \int_{0}^{T} dt \sum_{k} 
           \delta(t-t_{i}^{k}) \right>    
        =    \nu_p T \; ,
 \nonumber
\ee

\noindent
and its variance can be calculated using the autocorrelation
defined in eq. (\ref{autocorrelation}) as \citep{Renart+07}

\bea
 \sigma^2_{N,p}(T) &=& \left< \int_{0}^{T} dt \int_{0}^{T} dt'  \sum_{k,k'}
       \delta(t-t_{i}^{k})\delta(t'-t_{i}^{k'}) \right>
       - \left< N_p(T) \right>^2
 \nonumber
\\
 &=& \int_{0}^{T}dt \int_{0}^{T} dt \; C_p(t,t')
\nonumber
\\ 
 &=& \nu_p T  + \nu_p (F_p - 1) (T -\tau_c (1-e^{-T/\tau_c}))     \; .
 \label{eq:count-var-T}
\eea

\noindent
Therefore, the variance of the spike count grows linearly with $T$ for
long windows $T \gg \tau_c$, where it takes the value

\be
  \sigma^2_{N,p}(T)= F_p \nu_p T \;.
\label{eq:count-var}
\ee

\noindent
(see Fig.(\ref{fig:FN} C)).
Thus, fixing only the Fano factor in the autocorrelation function keeps fixed
the variance in the spike count for long $T$, as this variance is
independent of $\tau_c$.  Changing $\tau_c$ does not alter the total
spike count fluctuations, only the temporal precision in which they occur.
Notice that the inclusion of the Fano factor in the autocorrelation
function, eq.  (\ref{autocorrelation}), is consistent with its
definition for long $T$ in eq. (\ref{eq:F_NTdef}).  
Notice also from eq. (\ref{eq:count-var-T}), that the variance
of the spike count is $\nu_p T$ for short $T \ll \tau_c$, and therefore
the afferent spike train looks like a Poisson spike train when it is
sampled during brief time windows.  However, as soon as $T$ is
comparable with the correlation time, the variance of the spike count
starts to take into account the temporal correlations in the spike
train, and when $T$ becomes very large, all effects are included and
the variance is $F_p \nu_p T$, eq.  (\ref{eq:count-var}) 
(see Fig.\ref{fig:FN}).  
We will show below that, for the LIF neuron we are considering, 
whether the input is seen as having significant temporal correlations 
or not depends on how the timescale of these correlations compares
to the neuron's membrane time constant.

\subsection{Cross-correlograms}
\label{sec:cross-corr}

We have also considered the possibility that spikes in different trains
are correlated. 
When the activity of two neighbouring neurons is recorded,
the cross-correlogram computed from their discharges can sometimes present a
single peak with or without damped oscillations (e.g. \citep{Per+67,Aer+89,deC+96}). 
A prominent peak at zero time lag means
that the two neurons tend to fire synchronously, while if a dip is
observed, when one neuron fires the other is more likely to be silent.
Very often, the cross-correlograms
can be approximated by an exponential function (e.g. \citep{deC+96,Bair+01-tauc})
The cross-correlogram is therefore modeled here as an
exponential,

\begin{eqnarray}
C_{pq}(t,t')&\equiv& 
\left< \left( \sum_{k_p} \delta(t-t_{i}^{k_p}) -\nu_p  \right) 
\left( \sum_{k_q} \delta(t'-t_{j}^{k_p}) -\nu_q  \right) \right>
\nonumber 
\\
&=& \left< \sum_{k_p,k_q} \delta(t-t_{i}^{k_p})\delta(t'-t_{j}^{k_q}) \right> 
  -\nu_p \nu_q
\nonumber 
\\
&=& \sqrt{\nu_p \nu_q}\left(\frac{\rho_{pq} \;\sqrt{F_p\; F_q}}{2 \tau_{c}} \right)\; e^{-\frac{\mid t-t'\mid}{\tau_c}} 
\; ,
\label{crosscorrelation}
\end{eqnarray}

\noindent
where $C_{pq}(t,t')$ is the two-point correlation function between the
trains $(i,j)$ in populations $p$ and $q$ ($p,q=E,I$). This cross-correlation
function is illustrated in Fig.(\ref{fig:rho} B). As in the case
of the autocorrelation defined in eq. (\ref{autocorrelation}), the
two-point correlation function expresses the probability density of
finding a spike of a train in population $p$ at time $t$ along with a
spike of a train in population $q$ at time $t'$, 
from which the probability density of finding them by chance, $\nu_p
\nu_q$, is subtracted.  
The magnitude of the cross-correlations is determined by the
correlation coefficients $\rho_{pq}$ of the spike counts (see its
definition in eq. (\ref{eq:rho})). For the sake of simplicity, we take
all the correlations in the problem to have the same time constant
$\tau_c$.

To better understand the effects of cross-correlations on the input
statistics, we calculate the covariance between the count of spikes
emitted by the neuron $i$ from population $p$ and the count of spikes
emitted by the neuron $j$ from population $q$ as an integral of the
cross-correlation function, eq. (\ref{crosscorrelation}), as

\bea
 && \left< \left( N_p(T)-\left< N_p(T) \right> \right)
                  \left( N_q(T)-\left< N_q(T) \right> \right) \right> =
      \left<  N_p(T) N_q(T) \right> - \nu_p \nu_q T^2
 \nonumber
\\
 && \;\;\;\;\;\;\;\;= \left< \int_{0}^{T} dt \int_{0}^{T} dt'  \sum_{k_p,k_q}
       \delta(t-t_{i}^{k_p})\delta(t'-t_{j}^{k_q}) \right>
       - \nu_p \nu_q T^2
 \nonumber
\\
 && \;\;\;\;\;\;\;\;=  \int_{0}^{T} dt \int_{0}^{T} dt' \;  C_{pq}(t-t')
 \nonumber
\\
 && \;\;\;\;\;\;\;\;= \sqrt{\nu_p \nu_q} 
    \left( \rho_{pq} \sqrt{F_p\; F_q} \right)
    \left( T-  \tau_c(1-e^{-T/\tau_c})  \right)
  \; .
  \label{eq:cov-corr}
\eea

\noindent
This covariance measures the correlation in the spike count
fluctuations during a time $T$ from two presynaptic spike trains.  Notice that for
$T$ much shorter than the correlation time, this covariance is zero,
that is, the spike counts of the two neurons become independent. This
is true because for short $T$ the spike trains look like
uncorrelated Poisson trains. However, for time windows which are longer
than the correlation time, the covariance is non-zero and approaches a
linear behavior. This covariance as a function of the integration window 
is represented in Fig.(\ref{fig:rho} C).

The correlation coefficient is defined as the ratio of the covariance
and the product of the deviations in the spike counts of both neurons,
as

\be
 \rho_{pq}=\frac{\left< \left( N_p(T)-\left< N_p(T) \right> \right)
                  \left( N_q(T)-\left< N_q(T) \right> \right) \right> }
               {\sigma_{N_p}(T) \sigma_{N_q}(T)}  
 \label{eq:rho}
\ee

\noindent
for long $T$. Notice from eq. (\ref{eq:cov-corr}) 
that the inclusion of the correlation coefficient
in the cross-correlation, eq. (\ref{crosscorrelation}), is consistent
with the above definition. Changing the correlation time in the
cross-correlation, eq. (\ref{crosscorrelation}), changes its amplitude, 
but not the correlation coefficient between the two spike
trains. The Fano factors appear in eq. 
(\ref{crosscorrelation}) because the time integral of the
cross-correlation has to be zero if one of the trains does not have
spike count fluctuations ($F_N=0$).

\subsection{Writing the statistical properties of the total current}

The two-point correlation function of the total afferent current, eq.
(\ref{eq:current-deltas}), is defined as

\be
 C_{current}(t,t') \equiv \left< (I(t)-\left< I(t) \right>)
       (I(t')-\left< I(t') \right>) \right>   \; ,
  \label{eq:current_correlation_def}
\ee

\noindent
where the mean current $\left< I(t) \right>$ is calculated as in eq.(\ref{eq:mu}).
The correlation function should take 
into account both the auto- and cross-correlations of
the spike trains in the $E$ and $I$ populations given in eqs.
(\ref{autocorrelation}, \ref{crosscorrelation}). In Fig.
(\ref{fig:archi}) we depict a diagram with the correlations present in
the $E$ and $I$ neurons, whose spikes
trains impinge on the same target neuron. There are
$N_E$ excitatory neurons firing at rate $\nu_E$ and $N_I$ inhibitory
neurons with rate $\nu_I$. We assume that only a fraction $f_{EE}$
($f_{II}$) of the $N_E$ ($N_I$) excitatory (inhibitory) neurons are
correlated with other neurons within the same population, with a correlation coefficient $\rho_{EE}$
($\rho_{II}$). Also only a fraction $f_{EI}$ of the
excitatory neurons are correlated with a fraction $f_{EI}$ of the inhibitory
neurons, with a correlation coefficient $\rho_{EI}=\rho_{IE}$. 

Then, the correlation function of the current, eq.
(\ref{eq:current_correlation_def}), contains several contributions:

\begin{eqnarray}
 && C_{current}(t,t') = J_E^2 \; N_E \; C_E(t-t')+  J_I^2 \; N_I \; C_I(t-t') 
 \nonumber \\
       &&\;\;\;  +  J_E^2 \; f_{EE} \; N_E \; (f_{EE} N_E -1) \; C_{EE}(t-t')   
        +  J_I^2  \; f_{II}  \; N_I  \; (f_{II} N_I -1)  \; C_{II}(t-t') 
 \nonumber \\
       &&\;\;\; -  2  \; J_E  \; J_I  \; f_{EI}  \; f_{IE}  \; N_E  \; N_I  \; C_{EI}(t-t') \; .
 \label{eq:current_correlation_contributions}
\end{eqnarray}

\noindent
In this expression, 
the two first terms come from the auto-correlations of the spike trains in the
$E$ and $I$ populations. The third and fourth terms take into account the cross-correlation
between spike trains in the same $E$ or $I$ population. They are positive
because both $E$ and $I$ inputs contribute positively to enhance fluctuations.
The last term incorporates the cross-correlation between spike trains
one from the $E$ population and the other from the $I$ neuronal population, and
it is negative. Indeed, positive correlations between $E$ and $I$ neurons
always reduce synaptic fluctuations because arrival of an excitatory spike can be
cancelled out by arrival of another inhibitory spike, 
and this happens with higher than chance probability.  
Therefore, the effect of correlations within $E$ or $I$ neurons is always to increase
$C_{current}(t,t')$ in the direction of their cross-correlation functions,
 $C_{EE}(t-t')$ and $C_{II}(t-t')$, whereas the effect of correlations between
$E$ and $I$ spike trains is always to lower 
the current correlation function in an amount proportional
to $C_{EI}(t-t')$. 

Using the choices given in eqs. (\ref{autocorrelation},
\ref{crosscorrelation}), the two-point correlation function of the total
input current to the neuron can be written as

\be
 C_{current}(t,t') = \sigma_w^2 \left[ \delta(t-t') + 
 \frac{\alpha}{2 \tau_c} \; e^{- \frac{\mid t-t' \mid}{\tau_c}}
           \right] \;\;,
 \label{eq:current_correlation}
\ee

\noindent
where we call $\sigma_w^2$ the white noise variance, and $\alpha$ 
the {\em correlation magnitude}. They are expressed in terms of
the model parameters as

\begin{eqnarray}
\sigma_w^2 &=& J_E^2 \; N_E \; \nu_E + J_I^2 \; N_I \; \nu_I 
\nonumber \\
\alpha \;  \sigma_w^2 &=& J_E^2 \; \nu_E \; [(F_E - 1) + 
f_{EE } \; (f_{EE } \; N_E-1) \; F_E \; \rho_{EE}]
\nonumber \\
& & \mbox{} + J_I^2 \; N_I \; \nu_I \; [(F_I - 1) + 
f_{II} \; (f_{II } \; N_I-1) \; F_I \;\rho_{II}]
\nonumber \\ 
& & \mbox{} - 2 \; J_E \; J_I \; f_{EI} \; f_{IE} \; N_E \; N_I \;  
 \sqrt{\nu_E \; \nu_I } \; \sqrt{F_E \; F_I}  \; \rho_{EI}   \; .
\label{eq:variances}
\end{eqnarray}

\noindent
We define the total variance of the current, $\sigma^2_{eff}$, as the sum
of the white noise variance and the variance generated by
correlations, $\alpha \sigma_w^2$, that is,

\be
   \sigma^2_{eff}=\sigma_w^2(1+\alpha) 
   \label{eq:sigma_eff_alpha}
   \;. 
\ee

\noindent
The sign of the correlation magnitude determines the sign of the
correlations. If $\alpha > 0$, the current has {\em positive
correlations}, while if $\alpha < 0$, the current has {\em negative
correlations}. The minimum physically possible value for the 
correlation magnitude is $\alpha=-1$ $\;\;$
  \footnote{For large enough $T$
  ($T \gg \tau_c$), the variance of the integrated current, or accumulated
  charge $Q(t)=\int_{0}^{T} dt I(t)$, is calculated as
  
\be Var[Q(T)] =
  \int_{0}^{T} dt\int_{0}^{T} dt'\; C_{current}(t,t')=\sigma_{eff}^2 T 
  \; .
  \label{eq:var-Q-T} 
\ee

\noindent
Therefore, the variance of the current is just the
proportionality factor $\sigma_{eff}^2$. Notice that since
the variance of the current is non-negative, the correlation
magnitude has a lower bound at $\alpha=-1$. Lower values are
not physically possible because the variance of a 
real-valued stochastic variable cannot be negative.}.
If $\alpha = 0$, the current is uncorrelated. 
Notice that $\sigma^2_{eff}$ is very
sensitive to the fractions of correlated input trains, as these
fractions are multiplied by the number of connections from each
population to the square, which typically are of the order of $10^{3}-10^{4}$.
Also, from eq.  (\ref{eq:variances}) it is possible to see
that increasing the correlations between excitatory or inhibitory
neurons (either increasing $\rho_{EE}$ or $\rho_{II}$) enhances the
total variance, whereas correlations between excitatory-inhibitory
pairs ($\rho_{EI}$) always decrease it \citep{Sal+00}.

The parameters $\tau_c$ and $\alpha$ which appear in the definition
of the correlation function of the current, eq. (\ref{eq:current_correlation}),
fully characterize both the temporal
range and the intensity of the correlations relative to the white
noise variance $\sigma_w^2$. 
Although it is important to understand the effect 
of these two parameters on the neuronal firing response separately,
previous studies have not studied this problem. For instance, in
\cite{Fen+00} only the case $\tau_c=0$ is considered, which precludes
the characterization of the temporal scale of the correlations. On the
other hand, \citet{Sal+00} have changed simultaneously the values of
$\tau_c$ and $\alpha$ in their simulations.

\subsection{The sum of a large number of independent 
non-Poisson spike trains is not Poisson}

One point deserves clarification at this moment. It refers to the way many
simultaneous spike trains add up. The sum of many independent spike 
trains has been commonly approximated as a Poisson process (e.g. \citep{Daley+88,Ami+97}). 
Although this is in some cases a good approximation, it is worth emphasizing that
the sum of many independent point processes is not, in general, Poisson.
Indeed, the conditions for the sum-process to be truly Poisson 
are rather restricted (see e.g., \citet{Daley+88}). In particular, one
of the conditions implies that, on any time interval, only one event can be observed from
each individual point process. However, this is only expected to be a good approximation
for time windows much shorter than the typical inter-spike-interval of each neuron.
In general, a neuron will receive one, two or more spikes from the same
presynaptic neuron before it fires, not just at most one spike, as the Poisson
approximation strictly requires.

As expected from the rules of probability, 
adding up many independent spike trains results in 
a global spike train with an autocorrelation function 
which has exactly the same functional form
as those of the individual trains (note, however, that higher order properties are not 
necessarily conserved, i.e., the sum of many renewal processes may not be renewal). 
In particular, when $N$ independent spike trains 
with an autocorrelation $C(t,t')$ are added, the summed train has an autocorrelation $N
\times C(t,t')$ (\citep{Mor+02}, 
see also eq. (\ref{eq:current_correlation_contributions}) 
with $C_{EE(II,EI)}=0$ and $J_{E(I)}=1$). 
We further noted that even in the diffusion limit (i.e. $N
\rightarrow \infty$), when the individual firing rates $\nu$ are 
renormalized by $\nu/N$ to yield a finite two-point correlation function, the
auto-correlation function of the total input has exactly the same shape 
as the auto-correlation function of the individual spike trains. Later works have
also used this property \citep{Renart+07,Lindner06,Cateau+06-propag,Doiron+06},
which is relevant to describe the temporal aspects of correlations in networks
of spiking neurons.

Here we exemplify the above result using
the expression for the correlation function of the current,
eqs. (\ref{eq:current_correlation}-\ref{eq:variances}). It is easy to see
that the total current will show temporal correlations 
beyond the trivial delta function at zero time lag 
whenever $\alpha$ is different from zero and $\tau_c$ is not infinity. 
If the afferent spike trains are independent ($\rho=0$)
but they have exponential auto-correlations,
as those in eq. (\ref{autocorrelation}), then $\alpha$ will be different from
zero (see eq. (\ref{eq:variances})). This will happen for any choice of the number of
connections and synaptic strengths (different from zero). Therefore, no matter
which choices of the parameters are taken, the correlation-function of the total current 
can never correspond to a Poisson process with a larger rate, since an input Poisson
process will produce a correlation function equal to 
$C_{current}(t,t')=\sigma_w^2 \delta(t-t')$.
The above argument does not depend on the condition that the
correlations are exponential, but rather the same conclusion can be achieved from 
eq. (\ref{eq:current_correlation_contributions}) using any plausible autocorrelation
function $C_E(t,t')$ and $C_I(t,t')$ different from a delta function (i.e., different
from the autocorrelation function of a Poisson process).

\subsection{When the current can be approximated by a Gaussian current}

We have described the statistical properties of the total current, $I(t)$,
generated by correlated spike trains. However, the firing response of
a neuron receiving that current is not yet completely determined by the
mean and two-point correlation function of the current alone, 
eqs. (\ref{eq:current_correlation}-\ref{eq:variances}).
These quantities describe the statistical properties of a stationary current up to second
order, but higher order statistics in the input could also play a role
in shaping the firing response of the neuron. However, if the current $I(t)$
can be approximated by a Gaussian process, then, the current would be fully described
by its mean and two-point correlation function. In fact, Gaussianity 
naturally holds when the neuron is receiving a large barrage of
uncorrelated spikes per second each one inducing a membrane
depolarization $J$ very small compared to the distance between the
threshold and reset potentials, i.e., $J/(\Theta -H) \ll 1$ \citep{Ric77}.
When inputs are correlated, the net effect of correlations is to
increase effectively the size of the unitary depolarization (for
positive correlations), since two or more spikes are more likely to occur
together in time. We have estimated this renormalization in the
size of $J$ and determine that for the Gaussian approximation to be
valid with correlated input spike trains the condition 

\be
  \frac{J F}{(\Theta-H)}(1+f N \rho )\ll 1
  \label{eq:Gauss-condition}
\ee

\noindent
should hold. This is a heuristic formula, and it is explained qualitatively as
follows. The worst condition
in the presence of correlations occurs  when the correlation time $\tau_c$ is zero, that
is, when there is some chance that two or more spikes arrive at the same
time, increasing the effective size of each spike and worsening the Gaussian
approximation. 
One can estimate the mean number of spikes arriving
together to be $F \; (1+f N \rho )$, which
grows with the variability of the spike trains, the number of
correlated pairs and their correlation coefficient. 
As long as this number multiplied
by $J$ is small compared to $\Theta-H$, i.e., eq.(\ref{eq:Gauss-condition}), 
the Gaussian approximation is expected to be appropriate.
This indicates that if either
$F$, $fN$ or $\rho$ increases too much, the Gaussian limit will be broken.
When condition (\ref{eq:Gauss-condition}) is largely broken, 
as in \citep{Kuhn+03}, the Gaussian approximation is no longer valid. 
In particular in the limit of large $N$, it should hold that $\rho \sim 1/fN$,
so the correlation coefficients cannot remain finite as the size of the population 
of neurons with significant cross-correlations increases. 
If condition (\ref{eq:Gauss-condition}) is satisfied, 
the input current in our problem can be described 
as a Gaussian stochastic current fully defined in
terms of the mean $\mu= J_E N_E \nu_E - J_I N_I \nu_I$, the variance
$\sigma_w^2$, the correlation magnitude ($\alpha$) and
correlation time ($\tau_c$), as expressed in eq.
(\ref{eq:current_correlation}).

\subsection{Choosing the connectivity and correlation parameters}
\label{sec:parameters}

Because we are dealing with a model with many free parameters (see eq.
(\ref{eq:variances})), here we fix most  of them or make choices within a range
of realistic values. A single neuron receives typically $N_E \sim
5000-60000$ excitatory connections from other neurons
\citep{Cra67,DeF+92}. This accounts for $80$ per cent of the total
number of synapses; the remaining $20$ per cent corresponds to
inhibitory synapses \citep{Abe91}.  The dynamical range of cortical
neurons lies in the interval $\nu \sim 0-200Hz$ \citep{Alb93} although
lower rates are much more probable than higher ones \citep{Rol+98b}.
Synaptic strengths are between $J=0.1-1mV$ (\citep{Ami+97}; see the references therein).
Assuming a threshold of $20mV$ above the resting potential of the neuron, these unitary events
represent a fraction in the range $J \sim 5 \; 10^{-3}-10^{-2}$ of the total path to be
travelled from rest to firing threshold.

Fano factors of the spike count lying 
in the interval $1-1.5$ reveal higher irregularity in the
neuronal discharges than that expected from Poisson trains
\citep{Dea81,Sof+93,Sha+98,Alb93,Ste+98,Compte+03-F_N}.

The timescale of correlations varies from a few to several hundred
milliseconds, $\tau_c \sim 1-100 ms$ \citep{Tso+86,Goc+91}. 
For instance, in \citep{deC+96} the correlated activity of pairs of neurons in
primary auditory cortex in cats was recorded. The mean
half-width at half-height of the cross-correlograms peaks computed from
these pairs was $\sim 10 ms$, which corresponds to a correlation time
scale $\tau_c = 10 ms /ln2 \sim 15 ms$. 

\citet{Zoh+94} have reported correlation coefficients of $\rho=0.12$
between neighbouring cells in the middle temporal visual area (MT, or
V5). If any pair of neurons in a group of thousand units 
were correlated with such a magnitude and projected to a same target neuron, the
magnitude of the input fluctuations would be unrealistically large 
(see eq. (\ref{eq:variances})).  
In fact, the value $\rho=0.12$ only holds for units within local circuits, because
it is known that more distant neurons display much smaller 
correlation coefficients \citep{Lee+98-var}. Although a ``mean'' correlation
coefficient could have been considered \footnote{A mean correlation
  coefficient can be obtained by averaging the $\rho$ of each pair of
  neurons: $\left< \rho \right>=\int f(\rho) d \rho$.}, we have taken
into account the heterogeneity of pairwise correlations observed in the cortex
by assuming that only a fraction $f_{pq}$ of
neurons between populations $p$ and $q$ are indeed correlated with the
same correlation coefficient $\rho_{pq}$. This fraction could
represent the portion of presynaptic neurons located in the
surroundings of the target neuron, and thus embedded in the same local
circuits as this neuron, or a far neuronal population displaying
correlations between its units and projecting to the same target
neuron. To bound the effects of input correlations, 
we assume that around one per cent of the presynaptic 
neurons can be correlated. Such a small 
value of $f_{pq}$ still produces a large effect on the correlation magnitude
(see eq.  (\ref{eq:variances})), as will be also clear in
section (\ref{subsec:transient}).

The values of $\mu$, $\sigma^2_w$ and $\alpha$ therefore lie within rather
broad intervals. As an example of the typical values they can take, 
if a neuron receives $N_E=10^4$ excitatory connections, $N_I=2 \; 10^3$ inhibitory
connections, with synaptic strengths $J_E=5 \; 10^{-3}$ and $J_I=2 \; 10^{-2}$
(in units of the threshold), and they are
firing at $\nu_{E}=\nu_I=5Hz$, then $\mu=50Hz$ and $\sigma^2_w=5.3Hz$.
Assuming that there are only correlations between pairs of neurons
in the $E$ population ($\rho_{EI}=\rho_{II}=0$), being $f_{EE}=0.1$ the
fraction of those which are correlated, then $\alpha = 0.85$ if $F_E=F_I=1.5$
and the correlation coefficient is $\rho_{EE}=0.01$, or $\alpha = 4$ if $\rho_{EE}=0.1$.
When we present results from numerical simulations, the parameter 
values considered will be of the order of the ones mentioned above.


\section{Two ways of transforming the non-Markovian problem into a Markovian one}
\label{sec:FPequations}

As we have explained in the introduction, we aim at calculating the output
firing rate of a LIF neuron receiving a correlated input as
described in the previous sections.  
The main technical problem to study the response properties
of a neuron driven by correlated inputs analytically is that the stochastic process
defined by eq. (\ref{IF_equation}) with a current having correlations
as in eq.  (\ref{eq:current_correlation}) is non-Markovian, that is,
the time derivative of the membrane potential at each time depends on
the past history of the afferent current, not only on its present value.  
This fact complicates the solution of the problem. However, 
the process defined in eqs. (\ref{IF_equation},
\ref{eq:current_correlation}) can be expressed in a Markovian way by
generating the current $I(t)$ with the help of an Ornstein-Uhlenbeck
process \citep{Mor+02}. The stochastic current $I(t)$ generated in this
way displays exactly the same exponential correlations as eq.
(\ref{eq:current_correlation}).  This duplicates the number of
variables, but puts the problem in a suitable form (see eqs.
(\ref{current}, \ref{dynamic_z}) and (\ref{eq:I_alpha},
\ref{eq:z_alpha}) below).  We have found two different ways of
representing the correlated Gaussian current $I(t)$ satisfying eq.
(\ref{eq:current_correlation}). They only differ in the values of
$\alpha$ for which they hold.  While one of
them is more general because $\alpha$ can take any physical value
(including both positive and negative correlations), the other is simpler,
although it can be only used for $\alpha > 0$ (positive correlations).


\subsection{The first Representation for the dynamics of $I(t)$}
\label{subsec:pos_neg_corr}

The first representation of the current $I(t)$ that we discuss here
generates both positive ($\alpha > 0$) and negative ($\alpha < 0$)
correlations.  It has the form

\begin{eqnarray}
&&I(t) = \mu +  \sigma_w \eta(t)+ \sigma_w  \frac{\beta}{\sqrt{2\tau_c}}  z(t) 
\label{current} \\
&&\dot{z}(t) =  - \frac{z}{\tau_c} + \sqrt{\frac{2}{\tau_c}} \eta(t) \; , 
\label{dynamic_z}
\end{eqnarray}

\noindent
where $\eta(t)$ is a white noise random process with mean zero and
unit variance (i.e., $\left< \eta(t)  \right>=0$ and 
$\left< \eta(t) \eta(t') \right>=\delta(t-t')$), 
$\beta=\sqrt{1+\alpha}-1$ and $z(t)$ is an auxiliary
colored random process which obeys the Ornstein-Uhlenbeck process
(\ref{dynamic_z}) with the same white noise $\eta(t)$ (see, e.g., \citep{Ris89}).

It is easy to check that the current defined in eqs. (\ref{current},
\ref{dynamic_z}) generates a Gaussian waveform with mean $\left< I(t)
\right>= \mu$ and exponential correlations as in eq.
(\ref{eq:current_correlation}).  Defining $i(t)=(I(t)-\mu)/ \sigma_w $
we have

\bea
&&\left< i(t) \; i(t') \right> = 
\left< \left[\eta(t)+ \frac{\beta}{\sqrt{2\tau_c}} z(t) \right]   
\left[\eta(t')+ \frac{\beta}{\sqrt{2\tau_c}} z(t') \right] \right> 
  = \delta(t-t') +
\nonumber
\\
&& \;\;\;\; \frac{\beta}{\sqrt{2\tau_c}} \left< \eta(t) z(t') \right> 
+ \frac{\beta}{\sqrt{2\tau_c}} \left< \eta(t') z(t) \right> 
+ \frac{\beta^2}{2\tau_c} \left< z(t) z(t')  \right>   \; .
\label{pasos_corr}
\eea

\noindent
Assuming that $t'>t$ without loss of generality (because
$\left< i(t) \; i(t') \right>$ is symmetric in the steady state), the third term on the
right side of the eq. (\ref{pasos_corr}) vanishes. The second and
fourth terms are calculated using the solution of the stochastic equation, eq.
(\ref{dynamic_z}),

\be
 z(t) = \sqrt{\frac{2}{\tau_c}} \; e^{-t/\tau_c} 
\int_{0}^{t} ds \; e^{s/\tau_c} \; \eta(s) \; ,
\nonumber
\ee

\noindent
with the initial condition $z(0)=0$. We find that in the stationary
state ($t,t' \rightarrow \infty$, $t'-t=$ constant $> 0$)

\bea
 && \left< \eta(t) z(t') \right> \; 
     = \sqrt{\frac{2}{\tau_c}} \; e^{-(t'-t)/\tau_c} 
 \nonumber
 \\
 &&\left< z(t) z(t') \right> \; =  e^{-(t'-t)/\tau_c} \; .
 \nonumber
\eea

\noindent
Using these identities, the correlation function of the current $I(t)$ 
defined in eqs.(\ref{current}-\ref{dynamic_z}), denoted $C_{current}(t,t')$, 
can be written as

\begin{eqnarray}
C_{current}(t,t') &\equiv& \left< (I(t)-\mu)(I(t')-\mu) \right>  
\nonumber \\
        &=& \sigma_w^2 \left[\delta(t-t') + 
\frac{\beta(2+\beta)}{2 \tau_c} \; e^{- \frac{\mid t-t' \mid}{\tau_c}} 
\right]         \; ,
\label{two-point-corr}
\end{eqnarray}

\noindent
from where one sees that the correlation magnitude $\alpha$ is related
to the new parameter $\beta$ by $\alpha=\beta(2+\beta)$, 
an equation which has two independent
solutions, $\beta=\pm \sqrt{1+\alpha} -1$, both equally valid. We have
chosen $\beta=\sqrt{1+\alpha}-1$.  Remember that $\alpha$ has a lower
bound in $-1$, which is obtained with $\beta=-1$.  For each solution
there is a one-to-one mapping from $\alpha \in [-1,+\infty)$ to
$\beta$, and thus, all physically realizable 
positive and negative correlations are included in this formalism.


The joint process defined by eqs.
(\ref{IF_equation}, \ref{current}, \ref{dynamic_z}) is Markovian and
driven by white noise.  Thus, the problem of finding the output
firing rate can be formulated according to its associated stationary
Fokker-Planck equation (FPE) \citep{Ris89}. The system of
eqs. (\ref{IF_equation}, \ref{current}, \ref{dynamic_z}) can
be simplified by the linear transformation

\be
V=\mu \tau_m + \sigma_w \sqrt{\frac{\tau_m}{2}} x
\label{eq:cambio-V-x}
\ee

\noindent
to obtain the set of stochastic equations

\bea
 &&\dot{x}(t)=-\frac{x(t)}{\tau_m} + \sqrt{\frac{2}{\tau_m}} \eta(t) 
            + \frac{\beta}{\sqrt{ \tau_m \tau_c }} z(t)
\nonumber
\\
&&\dot{z}(t) =  - \frac{z}{\tau_c} + \sqrt{\frac{2}{\tau_c}} \eta(t) 
\; .
\nonumber
\eea

\noindent
The FPE associated to these two equations is derived in detail in
Appendix \ref{derivationFPE}, and is given by

\begin{equation}
 [L_x  + \frac{L_z}{k^2} 
           + \frac{2}{k} \frac{\partial}{\partial x}(
                \frac{\partial}{\partial z} -\frac{\beta z}{2})] 
 P_{\beta}(x,z)
 = - \tau_m \delta(x- \sqrt{2} \hat{H}) J_{\beta}(z)
 \; ,
 \label{FP-equation}
\end{equation}

\noindent
where the differential operator $L_u$ is defined as $L_u =
\frac{\partial }{\partial u} u + \frac{\partial^{2} }{\partial^{2}
  u}$, and $k \equiv \sqrt{ \tau_c / \tau_m }$.  
Besides, $\hat H = \frac{H-\mu \tau_m} { \sigma_w
  \sqrt{\tau_m}}$ and $\hat \Theta = \frac{\Theta-\mu \tau_m} {
  \sigma_w \sqrt{\tau_m}}$.  The true reset and threshold values in
the new variable $x$ are $\sqrt{2} \hat H $ and $\sqrt{2} \hat \Theta
$, respectively. The function $P_{\beta}(x,z)$ is the steady state
probability density of having the neuron in the state $(x,z)$.  Since
the problem cannot be solved exactly as in the one-dimensional
diffusion case (see e.g.  \citep{Ric77,Ris89}), we have used a
perturbative expansion of the FPE in powers of $k^{-1}=\sqrt{ \tau_m / \tau_c }$.

A key quantity is the escape probability density flux at fixed $z$,
$J_{\beta}(z)$. Associated to the FPE (\ref{FP-equation}) there is a
probability density vector flux $\vec J_{\beta}(x,z)$ defined at each point
on the plane $(x,z)$ (\cite{Ris89}, Chapter 6, pag. 133). 
It measures the direction and the
intensity of the probability density flux at each point $(x,z)$.  For
our FPE it has the expression

\be
 \vec J_{\beta} (x,z) =\frac{1}{\tau_m} [- \frac{\partial }{\partial x} - x 
 -\frac{1}{k} ( \frac{\partial }{\partial z} - \beta z)      
 \;\;  ,  \;\;
 - \frac{1}{k^2}(\frac{\partial }{\partial z} + z) 
 -\frac{1}{k} \frac{\partial }{\partial x}] P_{\beta}(x,z) \; .
\label{vec_J_beta}
\ee

\noindent
The probability density flux satisfies the so-called continuity equation

\be
 \vec \nabla  . \vec J_{\beta}(x,z) +   
 \tau_m \delta(x- \sqrt{2} \hat{H}) J_{\beta}(z) = 0 \; ,
 \label{divergencia1}
\ee

\noindent
where $\vec \nabla=[\frac{\partial }{\partial x},\frac{\partial
}{\partial z}]$ is the divergence operator. Eq. (\ref{divergencia1})
is equivalent to the FPE (\ref{FP-equation}), and expresses
the conservation of the total probability over time. The escape probability
density flux $J_{\beta}(z)$ is just the $x$-component of the probability
density flux (\ref{vec_J_beta}) evaluated at threshold:

\be
 J_{\beta}(z)=\frac{1}{\tau_m} \left(- \frac{\partial }{\partial x} - x 
 -\frac{1}{k} (\frac{\partial }{\partial z} - \beta z) \right)
 P_{\beta}(x,z) |_{x=\sqrt{2} \hat \Theta}    \; .
\label{J_x_beta}
\ee

\noindent
The escape probability density flux 
appears in eq. (\ref{FP-equation}) as a
source term representing the reset effect: whenever the potential $V$
reaches the threshold $\Theta$, it is reset to the value $H$ with the
same $z$ distribution that it had when it escaped.
This holds because the particular value of $z$ at the moment of
the generation of each spike has to be
conserved for the next inter-spike interval since, as opposed to $V$,
$z$ is not reset after an action potential. Crucially, this
self-consistency condition complicates the solution 
of the FPE (\ref{FP-equation}). The escape probability 
density flux in eq. (\ref{J_x_beta}) is exact if $\tau_{ref}=0$ (or approximately if
$\tau_c \gg \tau_{ref}$, because in this case the variable $z$ has slow dynamics
and therefore its probability distribution at a time $\tau_{ref}$ after the 
emission of an output spike is very similar to its distribution
at the moment of the spike).

Let us notice that this first representation of $I(t)$ can be used not
only for analytical calculations but also for the numerical generation
of exponentially correlated currents, as it is shown in Fig.
(\ref{fig:correlogram}), were we show the exponential two-point
correlation function of a current $I(t)$ generated numerically by eqs.
(\ref{current}, \ref{dynamic_z}) and that predicted by eq.
(\ref{two-point-corr}).  Additionally, in a later section this
representation will be employed in the numerical analysis of the
response of LIF neuron to negative and positive correlations 
(Sec \ref{sec:results}).


\subsection{The second Representation for the dynamics of $I(t)$}
\label{subsec:pos_corr}

\noindent
An afferent current $I(t)$ with non-negative exponential correlations
obeying eq. (\ref{eq:current_correlation}) 
can also be generated by the set of equations

\begin{eqnarray}
&& I(t)= \mu +  \sigma_w \eta(t)+ \sigma_w  
\sqrt{\frac{\alpha}{2\tau_c}}  y(t) 
\label{eq:I_alpha} \\
&&\dot {y}(t) = - \frac{y}{\tau_c} + \sqrt{\frac{2}{\tau_c}} \zeta(t) \; .
\label{eq:z_alpha}
\end{eqnarray}

\noindent
Here $\eta(t)$ and $\zeta(t)$ are two independent white noise
processes with mean zero and unit variance.  The two-point correlation
of $I(t)$ can be calculated as in eq. (\ref{pasos_corr}), with the exceptions that
$\beta$ in eq. (\ref{pasos_corr}) is replaced 
by $\sqrt{\alpha}$ and the two white noises are
not correlated. Then, only the terms analogous to the first and fourth
terms in eq.  (\ref{pasos_corr}) are nonzero. Because $\sqrt{\alpha}$
is a real number, the correlation magnitude $\alpha$ has to be positive in this
representation.

From the set of eqs.
(\ref{IF_equation}, \ref{eq:I_alpha}, \ref{eq:z_alpha}) and making the
linear transformation defined in eq. (\ref{eq:cambio-V-x}) we obtain
the FPE (the derivation is similar to the one presented 
in Appendix \ref{derivationFPE} )

\begin{equation}
\left[ L_x  + \frac{L_y}{k^2} 
           - \frac{\sqrt{\alpha} y}{k} \frac{\partial}{\partial x}
           \right] 
           P_{\alpha}(x,y)
= - \tau_m \delta(x- \sqrt{2} \hat{H}) J_{\alpha}(y)  \; .
\label{FP-equation2bis}
\end{equation}

\noindent
The linear differential operator $L_u$ has been defined as in 
Sec. \ref{subsec:pos_neg_corr}, and again $k \equiv \sqrt{ \tau_c / \tau_m }$. 
As in the previous representation, the escape probability density flux
$J_{\alpha}(y)$ acts as a source term injecting current at the reset
potential at the same rate and with the same distribution in $y$ as
when it escaped (here we have to assume that $\tau_{ref}=0$, or 
$\tau_c \gg \tau_{ref}=0$).  It represents the probability
current in the direction of $x$ evaluated at threshold. The
probability density vector flux for this FPE is

\be
\vec J_{\alpha} (x,y) =\frac{1}{\tau_m} [- \frac{\partial }{\partial x} - x 
-\frac{\sqrt{\alpha} y}{k}      
\;\;  ,  \;\;
-\frac{1}{k^2}( \frac{\partial }{\partial y} + y)]  P_{\alpha}(x,y) \; .
\label{vec_J_alpha}
\ee

\noindent
Its continuity equation is

\be
 \vec \nabla  . \vec J_{\alpha}(x,y) +   
 \tau_m \delta(x- \sqrt{2} \hat{H}) J_{\alpha}(y) = 0  \; ,
  \nonumber
\ee

\noindent
equivalent to the FPE (\ref{FP-equation2bis}),
and the escape probability density flux is defined as

\be
J_{\alpha}(y)=\frac{1}{\tau_m} \left(- \frac{\partial }{\partial x} - x 
-\frac{\sqrt{\alpha} y}{k} \right)  
P_{\alpha}(x,y) |_{x=\sqrt{2} \hat \Theta}  \; ,
\label{J_x_alpha}
\ee

\noindent
The FPE (\ref{FP-equation2bis}) will be useful for finding a perturbative
solution to the first passage time problem in powers of 
$k = \sqrt{ \tau_c / \tau_m }$, that is, for short $\tau_c$. We have found
this representation especially useful for this purpose, since this limit is
harder to obtain from the first representation.

\subsection{Conditions over the probability density distribution 
and probability density flux}
\label{subsec:self_cond}

For both representations of exponential correlations, the probability
density and the escape probability density flux must be determined
such that they obey the set of conditions:

\begin{enumerate}
\item Normalization of the probability density, 
 \be
 \tau_{ref}\nu_{out}+ 
 \int_{-\infty}^{\sqrt{2} \hat \Theta} dx \int_{-\infty}^{\infty} dw \; P_r(x,w)=1
 \label{condition1}
 \ee
\item Threshold vanishing condition, 
 \be
 P_r(\sqrt{2} \hat \Theta,w)=0
 \label{condition2} \ee
\item The output firing rate is given by 
 \be
 \nu_{out}=  \int_{-\infty}^{\infty} dw \; J_r(w).
 \label{condition3}
 \ee
\item The escape probability density flux has the form
 \be
 J_r(w)= -\frac{1}{\tau_m} \frac{\partial}{\partial x} 
           P_r(x,w)|_{x= \sqrt{2} \hat \Theta}
 \label{eq:escape_probability}
 \ee
\end{enumerate}

\noindent
where $r=\alpha,\beta$ is the representation label, and $w$ stands for
both $z$ and $y$. Condition (\ref{condition1}) is a normalization condition
stating that with probability $\tau_{ref} \nu_{out}$ the neuron is in
the refractory period. Condition (\ref{condition2}) states that at the firing
threshold, the probability density has to be zero (notice that the density can be
defined to be zero above threshold, so this condition is a continuity condition
at the threshold boundary). This is so because otherwise
the flux in eq. (\ref{eq:escape_probability}), which includes a derivative evaluated
at threshold, would be infinity.
The output firing rate of the neuron, $\nu_{out}$,
is obtained by integrating the escape probability density flux over $w$,
condition (\ref{condition3}). To write down $J_r(w)$ in condition
(\ref{eq:escape_probability}) we have used the condition
(\ref{condition2}) applied to eqs. (\ref{J_x_beta}) and (\ref{J_x_alpha}). 
Notice that precisely because of condition
(\ref{condition2}), the escape probability density flux, eq.
(\ref{eq:escape_probability}), has exactly the same expression in
both representations.

While solving the FPEs in both representations, it is usually
easier to employ the exact condition

\be
  \int_{-\infty}^{\sqrt{2} \hat \Theta} dx  \; P_r(x,w) 
   = (1-\nu_{out} \tau_{ref}) \; \frac{e^{ - w^2/2}}{\sqrt{2 \pi}}
 \;,
 \label{eq:Gauss-z-y}
\ee

\noindent
which is directly obtained from the equations for $z$ or $y$ (eqs.
(\ref{dynamic_z}) and (\ref{eq:z_alpha}) respectively) and the
condition that there is a fraction $\nu_{out} \tau_{ref}$ of neurons
in the refractory state. Eq. (\ref{eq:Gauss-z-y}) states that the
marginal distribution of $w$ is a normal distribution, as it corresponds
to the stationary distribution of an Ornstein-Uhlenbeck process
(eqs. (\ref{dynamic_z},\ref{eq:z_alpha})).
Notice that it is consistent with eq. (\ref{condition1}).


\section{Output firing rate for long and short $\tau_c$}
\label{sec:solution}

The next step is to compute the output firing rate using the FPEs.  We
found feasible to evaluate it from the first representation, eq.
(\ref{FP-equation}), for long correlation times ($\tau_c \gg \tau_m$),
and from the second representation, eq. (\ref{FP-equation2bis}), for both
short and long correlation times. In the two
cases we propose a perturbative expansion of the solution $P_r(x,w)$ in
powers of a representative temporal scale parameter (a convenient
power of $k \equiv \sqrt{ \tau_c / \tau_m }$).


\subsection{Long $\tau_c$ limit using the first representation}
\label{subsec:large_tauc}

In this limit we expand both the probability density and the escape probability
density flux as a series in powers of $k^{-1} = \sqrt{ \tau_m / \tau_c }$,

\bea
 && P_{\beta}(x,z)=h_0(x,z) + k^{-1} h_1(x,z) + k^{-2} h_2(x,z) + O(k^{-3})
 \label{eq:P_beta_expansion}
\\ 
 && J_{\beta}(z)=J_{0,\beta}(z) + k^{-1} J_{1,\beta}(z) +  
 k^{-2} J_{2,\beta}(z) + O(k^{-3})
 \; . 
 \label{eq:J_beta_expansion} 
\eea

\noindent
Each term $J_{i,\beta}$ in this expansion must satisfy the 
condition (\ref{eq:escape_probability}),  

\begin{equation}
J_{i,\beta}(z)= -\frac{1}{\tau_m} \frac{\partial}{\partial x} 
           h_{i}(x,z)|_{x= \sqrt{2} \hat \Theta}  \; .
\label{eq:escape_probability_order}
\end{equation}

\noindent
Let us proceed to the calculation by replacing the
expansions (\ref{eq:P_beta_expansion}, \ref{eq:J_beta_expansion}) into
the FPE (\ref{FP-equation}). This substitution generates a set of
equations for $P_{i,\beta}$ that can be solved consistently with
conditions (\ref{condition1} - \ref{condition3}).  The main steps of
the procedure are given in Appendix \ref{apendix_long}.

The resulting escape probability density flux $J_{\beta}(z)$ is found to be, up to
$O(k^{-2})$

\begin{eqnarray}
&&J_{\beta}(z) = \frac{e^{-z^2/2}}{\sqrt{2 \pi}} \; [ \nu_0 +
   \sqrt{\frac{\tau_m^3}{\tau_c}}
      \frac{(2 + \beta) \nu_0^2 (R(\hat \Theta) -
      R(\hat H))}{1-\nu_0 \tau_{ref}} \; z
\nonumber
\\
     && \;\;\;\;\;\;\;\;\;\;\;\;   + \frac{\alpha}{\tau_c} C  +
     \frac{\alpha C}{\beta^2 \tau_c (1-\nu_0 \tau_{ref})}
        (z^2-1) ]    \; ,
\nonumber 
\\
&&C  \equiv \tau_m^2 \nu_0^2 \;
\left[ \frac{\tau_m \nu_0 (R(\hat \Theta)-R(\hat H))^2}{1-\nu_0 \tau_{ref}} 
- \frac{\hat \Theta R(\hat \Theta) - \hat H R(\hat H)}{\sqrt{2}} \right] 
 \; . 
\nonumber
\end{eqnarray}

\noindent
Here $R(t)= \sqrt{\frac{\pi}{2}} \; e^{t^2}(1 + \rm{erf}(t))$, where
$\rm{erf}(t)=\frac{2}{\sqrt{\pi}} \int_{0}^{t} du \; e^{-u^2}$ is the
error function.  The rate $\nu_0$ is just the firing rate of a
LIF neuron driven by a white noise input with variance 
$\sigma^2_w$ \citep{Ric77}
 
\be
\nu_0^{-1}  =   \tau_{ref} +  \sqrt{\pi} \tau_m 
\int_{\hat H}^
{\hat \Theta} dt \; e^{t^2} (1 + \rm{erf}(t)) \;\; .
\label{nu0}
\ee

\noindent
Notice that $C$ is independent of $\tau_c$.  

We then use condition (\ref{condition3}) to find the output firing rate
valid for long $\tau_c$ and fixed $\alpha$

\be
\nu_{out} = \nu_0 + \frac{\alpha}{\tau_c} C\; .
\label{nu_out_big_tauc} 
\ee

\noindent
Several important conclusions can be extracted from this simple expression.
First, the effect of correlations is linear on $\alpha$ for long $\tau_c$.
That is, doubling $\alpha$ doubles the firing rate above the rate without
correlations, $\nu_0$. Notice also that $\alpha$ can be positive or negative, so
for negative correlations the effect on the rate is the opposite than for
positive correlations. Second, the firing rate of a LIF neuron with exponentially
correlated input approaches the firing rate in the absence of input correlations
as the correlation time increases.
This happens because, as the
correlation time becomes longer than the membrane time constant ($\tau_c \gg \tau_m$),
the neuron filters out the fluctuations provoked by input correlations. 
As a consequence, in the long $\tau_c$ limit and for finite correlation
magnitude, the correlated input to the neuron can be approximated by a white
noise process.  Therefore, in this limit, the observation of only the output
firing rate of the neuron does not allow to distinguish a correlated input
from other generated by the sum of many Poisson point processes in the
diffusion limit.
This result is important, as it determines when inputs
with complex correlation structure (i.e., with several correlation timescales)
can be approximated by white noise.


\subsection{Long $\tau_c$ limit using the second representation}
\label{subsec:large_tauc_second}

In this section we calculate the firing rate in the long $\tau_c$ limit using
the FPE (\ref{FP-equation2bis}).  Although the FPE (\ref{FP-equation2bis}) is
more restrictive than the FPE (\ref{FP-equation}) (it only describes positive
correlations, $\alpha>0$), it is analyzed here because it is much simpler and
can also be solved in the limit in which $\alpha/\tau_c$ is constant, i.e., for
arbitrarily large $\alpha$.
In fact, the FPE (\ref{FP-equation}) has been studied in
the limit in which $\alpha/\tau_c$ approaches to zero as $\tau_c$
rises, because the correlation magnitude was constant in that case. The real
advantage of using the second representation is that the predicted
firing rate is valid for larger values of $\alpha$, compared to the
formula (\ref{nu_out_big_tauc}).

We start from the FPE (\ref{FP-equation2bis}) and assume that the
factor $\sqrt{\alpha}/k$ is constant ($k \equiv \sqrt{ \tau_c / \tau_m }$). 
We thus define

\be
 \gamma=\frac{\sqrt{\alpha}}{k} \; .
 \nonumber
\ee

\noindent
Inserting this parameter in the FPE  (\ref{FP-equation2bis}) we obtain 

\begin{equation}
\left[ L_x  - \gamma y \frac{\partial}{\partial x} 
       + \frac{L_y}{k^2} \right] 
           P_{\alpha}(x,y)
= - \tau_m \delta(x- \sqrt{2} \hat{H}) J_{\alpha}(y)
 \; , 
\label{FP-equation-gamma}
\end{equation}

\noindent
where $J_{\alpha}(y)$ reads as in eq. (\ref{eq:escape_probability}). The
solution of the FPE (\ref{FP-equation-gamma})
along with the conditions (\ref{condition1}-\ref{eq:Gauss-z-y}) in the long
$\tau_c$ limit is found by expanding $P_{\alpha}(x,y)$ and the escape probability
density flux $J_{\alpha}(y)$ in powers of $k^{-2}$ while keeping $\gamma$ fixed
as

\bea
 P_{\alpha}(x,y)=r_0(x,y) + k^{-2} r_1(x,y) + O(k^{-4})
 \nonumber
 \\
 J_{\alpha}(y)=J_{\alpha,0}(y) +  k^{-2} J_{\alpha,1}(y) + O(k^{-4})
  \; .
 \label{eq:P-J}
\eea

\noindent
To obtain the coefficients $r_i(x,y)$ and $J_{\alpha,i}(y)$ we proceed
as in section (\ref{subsec:large_tauc}).  In particular, conditions
(\ref{condition1} - \ref{eq:escape_probability}) are imposed order by
order. The main steps of the calculation are given in Appendix
\ref{apendix_long_second}. The results are here summarized up to
order $k^0$. The density $P_{\alpha}(x,y)$ up to $O(k^0)$ is

\be
 P_{\alpha}(x,y)=\tau_m J_{\alpha}(y) \; e^{-\frac{(x-\gamma y)^2}{2}}
 \int_{x}^{\sqrt{2} \hat \Theta} 
   du \; e^{\frac{(u-\gamma y)^2}{2}} \; {\cal H}(u-\sqrt{2}\hat H)
  \; , 
 \nonumber
\ee

\noindent
(${\cal H}(t)=1$ if $t>0$ and it is zero otherwise)
where the escape probability density flux 
$J_{\alpha}(y)$ up to the same order is

\be
 J_{\alpha}(y) = \frac{1}{\sqrt{2 \pi} \tau_m} \; e^{-\frac{y^2}{2}}
    \left[ \int_{\sqrt{2} \hat H -\gamma y}^{\sqrt{2} \hat \Theta -\gamma y}  
       du \; e^{\frac{u^2}{2}} 
       \int_{-\infty}^{u} dv e^{-\frac{v^2}{2}} 
    \right]^{-1}  \; .
 \nonumber
\ee

\noindent
The output firing rate at leading order is obtained by integrating
$J_{\alpha}(y)$ over $y$ as

\be
 \nu_{out} = \frac{1}{\sqrt{2 \pi} \tau_m} \int_{-\infty}^{\infty} 
      dy \; e^{-\frac{y^2}{2}}
    \left[ \int_{\sqrt{2} \hat H -\gamma y}^{\sqrt{2} \hat \Theta -\gamma y}  
       du e^{\frac{u^2}{2}} 
       \int_{-\infty}^{u} dv e^{-\frac{v^2}{2}} 
    \right]^{-1}   \; .
 \label{nu_out_big_tauc-gamma}
\ee

\noindent
Notice that this formula has been derived for $\tau_{ref}=0$.  Notice
also that only the leading order $k^0$ has been calculated. This order,
however, gives a firing rate which is much more accurate than the
firing rate obtained using the first representation in the same limit,
eq. (\ref{nu_out_big_tauc}). This is true because the firing rate in
eq. (\ref{nu_out_big_tauc-gamma}) depends on $\gamma$, which is a
function of the parameters $\alpha$ and $k$ ($\gamma \equiv \sqrt{\alpha}/k$).  
If the zero-th order
firing rate in eq. (\ref{nu_out_big_tauc-gamma}) is expanded in powers of
$k^{-1}$ for fixed $\alpha$, the same firing rate in eq.
(\ref{nu_out_big_tauc}) is found when correlations are positive
(In particular, if $\alpha=0$, then $\gamma=0$ and $\nu_{out}$ equals $\nu_0$, i.e.,
the well-known expression for the firing rate of 
a LIF neuron driven by white noise \citep{Ric77}).
This means that eq. (\ref{nu_out_big_tauc-gamma}) is exact up to $O(k^{-2})$,
and therefore the corrections to the firing rate arising from the terms
$O(k^{-2})$ in the expansion (\ref{eq:P-J}) should vanish. This is indeed
the case, as shown in Appendix \ref{apendix_long_second}. In addition,
the higher other corrections found in the expansion of eq.
(\ref{nu_out_big_tauc-gamma}) improve the prediction provided by eq.
(\ref{nu_out_big_tauc}), especially when $\alpha$ is very large, or when
$\alpha / \tau_c $ is kept constant (i.e. $\gamma$ constant) in the long $\tau_c$ limit.

The firing rate in eq. (\ref{nu_out_big_tauc-gamma}) has a very simple interpretation.
Since $y$ is slow compared to the voltage dynamics ($\tau_c > \tau_m$),
the firing rate of a LIF neuron receiving correlated noise can be calculated by
multiplying the firing rate of the LIF neuron receiving 
a frozen current proportional to $y$ (plus mean $\mu$ and white noise
with amplitude $\sigma_w$; this corresponds
to the function into the square brackets, divided by $\tau_m$ \citep{Ric77}
\footnote{Note that the function within the brackets can be expressed
in terms of the error function, similarly to eq. (\ref{nu0}) }) 
and the probability density of having
the value $y$, which in this case is a normal distribution because $y$ obeys
an Ornstein-Uhlenbeck process, eq (\ref{eq:z_alpha}). 
This expression, obtained here to describe the
effect of exponentially correlated inputs, has also been used to describe the effects of
synaptic filtering with both fast and slow linear synapses on the output firing
rate of a LIF neuron \citep{Mor+04}.

One important feature of the firing rate in the long $\tau_c$ limit,
eq. (\ref{nu_out_big_tauc-gamma}), is that it does not change as $\gamma$ is
kept fixed, that is, as the ratio $\alpha / \tau_c$ is kept constant. This
means that to obtain the same output firing rate for a longer correlation time, 
one has to increase proportionally the correlation magnitude so that
the loss of temporal precision is counterbalanced by an increase
in the excess of synchronous afferent spikes. This suggests a proportionality
law that could be tested experimentally using {\em in vitro} current injections
in which both the magnitude and the temporal precision can be controlled independently
(using e.g. eqs. (\ref{dynamic_z})).
Furthermore, for non-exponential correlation functions (e.g. oscillatory), 
it might be possible to define an effective correlation magnitude  
and an effective correlation time so that the output firing rate of the neuron
will not depend on them individually, but on their ratio.


\subsection{Short $\tau_c$ limit using the second representation}
\label{subsec:small_tauc}

In the regime of short $\tau_c$ the FPE (\ref{FP-equation2bis}) is
employed to find the output firing rate. Although the set of eqs.
(\ref{IF_equation}, \ref{eq:I_alpha}, \ref{eq:z_alpha}) only generates
positive exponential correlations, we use them because its associated
FPE can be solved perturbatively in powers of $k=\sqrt{\tau_c/\tau_m}$
and $\sqrt{\alpha}$. We have found the FPE (\ref{FP-equation}) including both
positive and negative correlations too involved to be studied
in the small $k$ limit.  Although the firing rate computed in this limit
from the FPE (\ref{FP-equation2bis}) is derived only for positive correlations, 
when the same formula
is employed for negative correlations, one finds an excellent
agreement with the numerical results.  This fact suggests that the
analytical continuation of our formula to negative correlations
($\alpha<0$) could match the true expression for that case.

Even when using the FPE (\ref{FP-equation2bis}), valid for positive
correlations, the short $\tau_c$ expansion is not easy to obtain. This
is because of the self-consistency condition
(\ref{eq:escape_probability}), which is hard to deal with. However, if
the correlation time $\tau_c$ is short compared to the refractory time
$\tau_{ref}$ ($\tau_{ref} \gg \tau_c $), the escape probability
density flux can be written as \citep{Doe+87}

\be 
J_{\alpha}(y)= \nu_{out} \; \frac{e^{-y^2/2}}{\sqrt{2 \pi}}   \; ,
\ee

\noindent
which solves automatically the conditions (\ref{condition3} -
\ref{eq:escape_probability}).  This approximation is good because
after a spike the variable $y$ approaches its Gaussian stationary
distribution in a time $\tau_c$, which we are taking shorter than
$\tau_{ref}$.

We now look for a solution of the FPE (\ref{FP-equation2bis}) 
of the form

\be
P_{\alpha}(x,y)=f_0(x,y)+ k f_1(x,y) + O(k^2) \;,
\label{P_alpha_expansion}
\ee

\noindent
and at the same time we expand the escape probability density flux
$J_{\alpha}$ in powers of $k$ or, equivalently, the unknown output
firing rate as

\be
\nu_{out}=\nu_{eff}+ k \nu_1 + O(k^2) \;\; .
\label{nu_alpha_expansion}
\ee

It can be shown that the solution $f_1(x,y)$ obtained from the
perturbative expansion does not satisfy the vanishing boundary condition
(\ref{condition2}) (see Appendix \ref{appendix:short}).  
To address this problem, we extend the formalism described in
\citep{Doe+87} to solve the short $\tau_c$ limit. As in
\citep{Doe+87}, our problem does not have a perturbative solution for
short $\tau_c$, and it is necessary to solve a boundary layer
problem. Details of these calculations are given in Appendix
\ref{appendix:short}. Briefly, the solution
$P_{\alpha}^{total}(x,y)$ is obtained as the sum of the perturbative
and an additional solution, valid close to the threshold, that we call
{\it boundary solution} $f_1^b(x,y)$:

\be
  P^{total}_{\alpha}(x,y)=f_0(x,y)+ k [f_1(x,y) + f_1^b(x,y) ] + O(k^2)
  \; .
  \label{P_alpha_total}
\ee

\noindent
It is now possible to satisfy the condition (\ref{condition2}) up to
order $k$.   Finally, the firing rate up to order $k$ can be calculated using
the condition (\ref{condition1}), resulting

\begin{equation}
\nu_{out}= \nu_{eff}(\alpha) - \alpha \sqrt{\tau_c \tau_m}  
                \nu_0^2  R(\hat \Theta) \;,
\label{nu_out_small_tauc}
\end{equation}

\noindent
where $\nu_0$ is defined as in eq. (\ref{nu0}) and

\be
\nu_{eff}^{-1}(\alpha)  =  \tau_{ref} + \sqrt{\pi} \tau_m 
		\int_{\hat H_{eff}}^
		{\hat \Theta_{eff}} dt \; e^{t^2} (1 + \rm{erf}(t))  \; .
\label{eq:nu_tauc0}  
\ee

\noindent
The effective reset and threshold potentials are defined as $\hat
\Theta_{eff}=\frac{\Theta-\mu\tau_m}{\sigma_{eff} \sqrt{\tau_m}}$ and
$\hat H_{eff}= \frac{H-\mu \tau_m}{\sigma_{eff} \sqrt{\tau_m}}$.
An important implication of eq. (\ref{nu_out_small_tauc}) is
that when $\tau_c=0$ the output rate is $\nu_{eff}(\alpha)$, 
equivalent to that of a LIF neuron receiving an 
{\it uncorrelated} input (white noise) with an effective signal variance

\be
 \sigma_{eff}^2=\sigma_w^2(1+\alpha)  \; .
 \label{eq:sigma_eff}
\ee

\noindent
In this case, the solution is {\em exact} for all $\alpha$. When $\tau_c
\neq 0$, it is correct only for small values of both $k$ and $\alpha >
0$. This expression indicates that when the correlation time is
zero, the effect of the input correlations is to increase the white noise variance
by a factor equal to $1+\alpha$.

A general expression for the firing rate of any IF neuron is presented in  
Appendix \ref{appendix:short-generic}. 
Again, for $\tau_c=0$, the firing rate is that of the IF neuron with input white
noise but with a renormalized variance as in eq. (\ref{eq:sigma_eff}). From this
maximum firing rate at optimal synchronization, the firing rate decreases
as $-\sqrt{ \tau_c }$ for fixed $\alpha$ (eq. (\ref{rates_teoricos.eq}),
analogous to eq. (\ref{nu_out_small_tauc}) for a LIF neuron), showing
that this large sensitivity to variations in the correlation time of the inputs
is a general property of IF neurons. 

\section{Summary of the analytical results}
\label{subsec:summary}

The analytical results, obtained in the first and second representations
of the current, and in the short and long $\tau_c$ limits, along with the
conditions upon which they are valid, are summarized in Table
(\ref{table1}).

\begin{table}
\centering
\begin{tabular}{|l|l|l|} \hline
                    &   &   \\ 
                   &   {\em First Representation}  & {\em Second Representation}  \\  
					 &   &   \\  \hline

 &   &   \\ 
 Short $\tau_c$   &   Not found  
                   &   $
							\nu_{out}= \nu_{eff}(\alpha) - \alpha \sqrt{\tau_c \tau_m}  
							\nu_0^2  R(\hat \Theta) \;,
						  $
					   
				   eq. (\ref{nu_out_small_tauc}). \\  
				 &   & \\
				 &	   & Valid for small and positive $\alpha$, and     \\  
				 &	   & {\em exact} for all positive $\alpha$ when $\tau_c=0$.      \\  
             &   &   \\  \hline 
              
  &   & \\             
  Long $\tau_c$    &  
                   $	\nu_{out} = \nu_0 + \frac{\alpha}{\tau_c} C\;, $
                       
                   &
                   $
					\nu_{out} = \frac{1}{\sqrt{2 \pi} \tau_m} \int_{-\infty}^{\infty} 
							dy \; e^{-\frac{y^2}{2}}
					$
					 \\  
					
					& eq. (\ref{nu_out_big_tauc}). 
					& $
					    \;\;\;\;\;\;
							\left[ \int_{\sqrt{2} \hat H -\gamma y}^{\sqrt{2} \hat \Theta -\gamma y}  
							du e^{\frac{u^2}{2}} 
							\int_{-\infty}^{u} dv e^{-\frac{v^2}{2}} 
							\right]^{-1}   \; ,
					$
					\\  
					& 
					& with $\gamma=\sqrt{ \alpha \tau_m / \tau_c   }$, 
						eq. (\ref{nu_out_big_tauc-gamma}). 
					
					\\  
					& Valid for small $\alpha$,
					&  
					
					\\  
					& positive and negative.
					& Valid for all (even large) positive $\alpha$. \\
    &   & \\    \hline
\end{tabular}
\caption{\label{table1} Analytical expressions for the output
firing rate of a LIF neuron receiving exponentially correlated inputs
with magnitude $\alpha$ and correlation timescale $\tau_c$
for the two representations of the current in both the short and long $\tau_c$ limits.}
\end{table}

The second representation allows one to calculate the firing rate for both
short and long $\tau_c$, while the first representation only allows the calculation
of the firing rate for long $\tau_c$. 
In \citep{Mor+02} we used the second representation to obtain the firing rate for
short $\tau_c$, and therefore the expression shown here is the same at that found
there. On the other hand, using the second representation for long $\tau_c$, here we
have been able to find a new expression for the firing rate,
eq. (\ref{nu_out_big_tauc-gamma}), which can be
applied for arbitrarily large $\alpha$, while that found in \citep{Mor+02}
using the first representation, eq. (\ref{nu_out_big_tauc}), could only
be applied for small $\alpha$.
The expressions valid for long $\tau_c$,
eq. (\ref{nu_out_big_tauc}) and eq. (\ref{nu_out_big_tauc-gamma}), are in fact
equivalent when the limit $\alpha \rightarrow 0$ is taken 
(Appendix \ref{apendix_long_second}). Note, however, that 
eq. (\ref{nu_out_big_tauc}) can be employed for negative $\alpha$, while
eq. (\ref{nu_out_big_tauc-gamma}) can only be used for positive $\alpha$.


\section{The effect of correlations on the firing response of spiking neurons}
\label{sec:results}
 
In this section we take advantage of the machinery developed in the previous sections. 
First, the prediction of the firing rate as a function of the
timescale and magnitude of input correlations is used to study
the role of synchrony on the stationary firing response of a LIF
neuron. Second, we study the firing response to
modifications of the correlation magnitude. Numerical solutions of the voltage
and noise equations to generate exponentially correlated noise are employed in
this case.

\subsection{Stationary firing response}
\label{subsec:results}

Although we have calculated the output firing rate both in the limit
$\tau_c \ll \tau_m$ and in the limit $\tau_c \gg \tau_m$,
before the effect of $\tau_c$ and $\alpha$ on the firing rate is described, 
we develop an interpolation procedure that allows us to use a single 
expression for all values of $\tau_c$. 
The interpolating curves have been determined by setting the firing
rate in the short correlation time range ($\tau_c < \tau_m$) as

\be
 \nu_{out}=\nu_{eff}+A_1 \sqrt{\tau_c}+ A_2 \; \tau_c \; ,
 \label{eq:nu_short_inter}
\ee

\noindent
where $A_1$ and $A_2$ are unknown functions of $\alpha$ and of the
neuron and input parameters, while in the long correlation time limit
($\tau_c > \tau_m$) the expression given in eq.
(\ref{nu_out_big_tauc}),

\be
 \nu_{out}=\nu_0+\alpha C/\tau_c \;, 
 \label{eq:nu_long_inter}
\ee

\noindent
was used. The functions $A_1$ and $A_2$ are determined by
interpolating these two expressions with conditions of continuity and
differentiability at a convenient interpolation point $\tau_{c,inter} \sim
\tau_m$. Although we have calculated analytically the function $A_1$
(eq. (\ref{nu_out_small_tauc})) for small $\alpha$, this procedure
takes into account higher order corrections which match more
accurately the observed data for larger values of $\alpha$, as those
used in some of our simulations (see below).
Therefore, eqs. (\ref{eq:nu_short_inter}-\ref{eq:nu_long_inter}) provide
an analytical formula for the output firing rate of a LIF neuron receiving
exponentially correlated input which is valid for all $\tau_c$.

We have performed numerical simulations of a LIF neuron driven by
Gaussian exponentially correlated input using eqs. (\ref{IF_equation},
\ref{current}, \ref{dynamic_z}). We use them to 
check the analytical results given in eqs.
(\ref{nu_out_big_tauc}, \ref{nu_out_big_tauc-gamma},
\ref{nu_out_small_tauc}) and validate
the interpolation made between the regimes of short and long $\tau_c$,
provided by eqs. (\ref{eq:nu_short_inter}-\ref{eq:nu_long_inter}).
When positive correlations are
considered ($\alpha > 0$), the interpolation procedure is robust against
changes in $\mu$ and $\sigma_w^2$. Crucially, the interpolating point
$\tau_{c,inter} \sim \tau_m$ does not vary too much, so that it can be
maintained approximately fixed for all input parameters.  For negative
correlations we have found more convenient to add to the expansion in eq.
(\ref{eq:nu_long_inter}) an extra term:
$\nu_{out}=\nu_0+\alpha \; C/\tau_c+B_1/\tau_c^2$. Then, this 
expression is made to match at
$\tau_{c,inter} \sim \tau_m$ the short $\tau_c$ regime given by the equation
$\nu_{out}=\nu_{eff}+B_2 \sqrt{\tau_c}$.

This interpolation is compared with simulation results in
Fig. (\ref{fig:grafica1}), providing good fits. The firing
rate increases as $\tau_c$ decreases (at fixed positive $\alpha$).
This corresponds to the intuitive result that positive correlations
between the presynaptic events produce a larger enhancement in the
output firing rate as the temporal window over which they occur
decreases.  On the other hand, when negative correlations are present
in the input, the effect of $\tau_c$ is reversed: the firing rate
increases as $\tau_c$ increases. Negative correlations produce a
deficit in current fluctuations that decreases the firing rate.  This deficit
is not noticeable if $\tau_c$ is very long compared with $\tau_m$.
These results show that correlations with {\em fixed} magnitude
$\alpha$ have different effects on a target neuron
depending on the value of their correlation timescale. Correlations
are not perceived by neurons if the temporal precision they occur is
larger than the membrane time constant of those neurons.  As it can
be appreciated in Fig. (\ref{fig:grafica1}), when $\tau_c$ is of the order of
$40ms$ (twice longer than $\tau_m$) the output firing rate of
the neuron approaches the firing rate obtained by an input
without correlations ($\alpha=0$, dashed-dotted line). Only if $\tau_c <
\tau_m = 20ms$, the presence of correlations is noticeable. 
As noted above, this implies that, 
from the point of view of the output firing rate, correlations in the input
can be neglected, i.e., a white-noise input description is appropriate, 
when $\tau_c$ is significantly longer than $\tau_m$ (note, however, that
there is not an absolute value of $\tau_c$ for which correlations
can be neglected, rather, this value will increase with $\alpha$).
 
In Fig. (\ref{fig:grafica-Bal-NoBal}) we use the predictions of
eqs. (\ref{nu_out_big_tauc},\ref{nu_out_big_tauc-gamma}) valid for
long $\tau_c$. Here, large values of correlation magnitude, $\alpha$, are used.
The predictions are compared with simulations of neurons in the
subthreshold (left) and the suprathreshold (right) regimes.
The subthreshold and suprathreshold regimes are defined by $\mu \tau_m <
\Theta$ and $\mu \tau_m > \Theta$ respectively, and they correspond
to the fluctuation and drift dominated regimes. 
The prediction by eq. (\ref{nu_out_big_tauc-gamma}) is very good
even for intermediate $\tau_c \sim \tau_m$ in both regimes.
In contrast, the firing rate for long $\tau_c$
given in eq. (\ref{nu_out_big_tauc}), provides poorer fits
(dotted line in the left panel) in the subthreshold regime, and even
poorer in the suprathreshold regime when very large values of $\alpha$ are
used (not shown). This is because the second prediction of the firing rate
was obtained for fixed $\alpha$.

The figure also shows that the effect
of correlations is quite different for a neuron receiving subthreshold
or suprathreshold inputs.  For subthreshold inputs, positive
correlations always increase the firing rate relative to the case
without correlations, and the firing rate decreases as the timescale
of correlations becomes broader.  However, for suprathreshold inputs a
different qualitative behavior is observed, at least for small white noise
variances.  Positive correlations with long enough $\tau_c$ give an
output firing rate smaller than the basal rate without correlations,
although this effect is very small (notice the large values of
$\alpha$ that have been used). A minimum firing rate is attained when
the correlation timescale is longer than the membrane time constant of
the neuron, and the exact value of $\tau_c$ at which the minimum
occurs is roughly predicted by the analytical formula
(\ref{nu_out_big_tauc-gamma}). When the white
noise variances become larger, this counterintuitive effect of
correlations disappears, and the profile is much more similar to the
subthreshold case, but with much smaller correlation-induced changes.

Overall, this analysis shows that neurons are more
sensitive to correlations in the subthreshold than in the suprathreshold 
regime, what is not surprising, since in the first regime
spiking is driven by input fluctuations and correlations
enhance them \citep{Mor+02,Sal+01a}.


\subsection{Transient firing response}
\label{subsec:transient}

Another important question is how fast a 
neuron can respond to {\em pure} changes in the
correlation magnitude $\alpha$, that is, when both the afferent mean current
and white noise variance $\sigma_w^2$ are fixed. In our work \citep{Mor+02}
we have shown that changes in correlation magnitude can be transmitted
very fast by the firing rate of spiking neurons 
even when the timescale of those correlations
is quite large. Those firing responses are also compared here with the response
to sudden jumps in mean input current.

Let us write the instantaneous firing rate for the time dependent FPE, either
in the first or in the second representation, as 
(see eq. (\ref{eq:escape_probability}))

\begin{equation}
\nu_{out}(t) = - \frac{\sigma_w^2 (t)}{2} \frac{\partial
}{\partial V} \int_{-\infty}^{\infty} dw P(V,w,t)|_{V=\Theta}
 \; .
\label{eq:rate_t}
\end{equation} 

\noindent 
For the sake of clarity, we have come back to the physical quantity $V$
and used its distribution $P(V,w,t)$ ($w=z,y$). A similar equation for
the instantaneous firing rate of a one-dimensional FPE has been used
by \cite{Silberberg+04} to predict that any instantaneous modification
in the white noise variance, $\sigma_w^2(t)$, produces an immediate
change in the output firing rate of the neuron. Besides, as we have
shown before, the exact form of eq. (\ref{eq:rate_t})
for $\tau_c=0$ corresponds to a neuron receiving (uncorrelated) input white noise 
with effective variance $\sigma_{eff}^2=\sigma_w^2 (1+\alpha)$,
eq. (\ref{eq:sigma_eff_alpha}). This gives \citep{Mor+02}

\be
 \nu_{out}(t) = - \frac{\sigma_{eff}^2
 (t)}{2} \frac{\partial}{\partial V} \int dw P(V,w,t)|_{V=\Theta}
  \; .
 \nonumber
\ee

\noindent
Now it is clear that any change in $\alpha$ will produce an
immediate change in $\nu_{out}(t)$, because the distribution
$P(V,t)= \int dw P(V,w,t)$ can only experience a smooth change (notice that the
trajectories generated by the equations for $V$ (e.g, see eqs.
(\ref{IF_equation}, \ref{current}, \ref{dynamic_z})) are continuous
under changes in $\alpha$).  This means that when
$\tau_c=0$, changes in the correlation magnitude ($\alpha$) 
will be felt immediately by the firing
response of the neuron. By analyticity arguments, the response under
changes in $\alpha$ will be also fast for non-zero $\tau_c$.

These predictions have been tested with numerical simulations, whose
results are shown in Fig. (\ref{fig:histograma1}). Initially the input
statistics is white noise, and some time later either the mean current
$\mu$ (bottom curve), or the white noise variance $\sigma_w$ (upper
curve), or the correlation amplitude $\alpha$ (two intermediate
curves) are changed independently. 
Changing abruptly the mean current only produces a slow
response with a timescale of the order of the membrane time constant.
However, in the absence of correlations, the firing rate changes
instantaneously under a sudden modification in the variance of the injected
current ($\sigma^2_w$). In agreement with our prediction, for short
$\tau_c$ the response is also very quick when the correlation changes
from $\alpha=0$ to a positive value. To quantify how fast the response
is, we computed the time $t_{cross}$ at which the instantaneous rate
reaches for the first time the value of the final stationary firing
rate. The inset in Fig.  (\ref{fig:histograma1}) shows that, as a
function of $\tau_c$, $t_{cross}$ initially grows but it soon
saturates at about $3ms$, even when $\tau_c$ is several hundred
milliseconds long. Thus, the correlation time is not a limiting factor
for fast transmission of information contained in correlation changes.
This result shows that information carried by correlated input
patterns can be transmitted with a timescale that is not limited by
the membrane time constant, what is not the case for signals embedded in the
mean input current \citep{Mor+02}. In \citep{Rud+01} the authors show that
correlation changes can be followed very rapidly by a spiking neuron.
Because they consider the case of perfect synchrony,
$\tau_c=0$, their conclusions are similar to those by
\citep{Silberberg+04}, because the case $\tau_c=0$ corresponds to a
simple renormalization of the current variance ($\sigma_w^2$), as we
have explained before (see eq. (\ref{eq:sigma_eff_alpha})).

These results show that  fast information transmission in cortex using 
spike correlations is theoretically possible.
As we have shown, changing the
mean afferent current produces slow responses if the neuron is in the
subthreshold regime, because the mean current has to be integrated in
a timescale $\tau_m$.  However, because of their fast transmission rate, 
correlation modulations can be an ideal candidate for transmitting information
rapidly. The fact that changes in $\mu$ do not evoke rapid responses does not
mean that rate codes are inefficient for transmitting information
rapidly. Rather, changes in the firing rate of "noisy" input spike trains
(e.g., as in a Poisson train) involve both changes in $\mu$ and in
fluctuations $\sigma_w$ \citep{Ric77}
and indeed also in $\alpha$ (see their definitions in eqs. (\ref{eq:variances})).  
Such white noise variance
and correlation magnitude modulations can be transmitted very fast,
while the mean current modulations produce a slower response.
Therefore, an increase in the firing rate of an irregularly
spiking presynaptic population will produce an output rate change
which contains information in at least two different timescales (one
short and another slow).


\section{Discussion}
\label{sec:discussion}

In this paper we have provided and thoroughly analyzed a theoretical 
framework to understand how 
temporal correlations affect the output firing response of neurons.
The main qualitative results we found are

\bi{
  
\item The neuron's output rate is very sensitive to precisely
  synchronized inputs with $\tau_c < \tau_m$.
  
\item The response decreases (increases) with the timescale $\tau_c$
  for positive (negative) correlations, and increases (decreases) with
  their magnitude $\alpha$.
  
\item The neuron response to sudden changes in the size of the
  correlations is very fast, regardless of the magnitude of the
  change and on the correlation time.  

} 
\ei

An important question is how our results can be incorporated into the modeling
of neural networks. Temporal and spatial correlations
are presumably relevant to correctly describe the dynamics
of realistic recurrent neuronal networks.
Recently in \citep{Renart+07} we have proposed an extended mean-field
approach to determine the firing rate and spiking variability of a large
network of LIF neurons.
In the classical mean-field theory, the neurons in the network are assumed
to fire in a Poisson and independent manner \citep{Ami+97-spont,Renart+03book},
so that the only free dynamical parameter in the dynamics of an homogeneous
population of neurons is its population firing rate. 
Our extension goes beyond the classical mean field theory by adding as a free parameter the spiking variability
of the network, that is, the coefficient of variation of the inter-spike-intervals,
$CV$. Then, the firing rate as well as the variability of the network can be
studied without the assumption that the spike trains are Poisson, corresponding
to the particular case $CV=1$. In particular, stationary states with
$CV>1$ would correspond to states of high spiking variability, while stationary
states of the network with $CV<1$, would correspond to more regular spiking
regimes of the neuronal dynamics.
The formalism presented in \citep{Renart+07} is based on the result 
that when the correlation time of the spike trains is short enough
($\tau_c \ll \tau_m$), then the input variability can be expressed as 
(see eqs. (\ref{eq:sigma_eff},\ref{eq:variances}); \citep{Mor+02})

\be
  \sigma_{eff}^2 = J_E^2 \; N_E \; CV_E^2 \; \nu_E + J_I^2 \; N_I \; CV_I^2 \; \nu_I 
  \label{eq:sigma_renor}
  \; ,
\ee

\noindent
assuming that there is no cross-correlations ($\rho = 0$). Since the output 
firing rate and the output $CV$ of
an integrate-and-fire neuron can be calculated exactly when the input is white noise 
\citep{Ric77}, then a mapping between the input rates and $CV$, and the output
rates and $CV$ can be constructed as

\bea
 \nu_{out} &=& f_{\nu}( \nu_{in}, CV_{in}  )
 \nonumber
\\
 CV_{out} &=& f_{CV} ( \nu_{in}, CV_{in}  )
 \label{eq:nu_out-CV_out}
 \; ,
\eea 

\noindent
where the functions $f_{\nu}$ and $f_{CV}$ are the expressions for the output
firing rate and $CV$ of the IF neuron receiving white noise input.
These equations define an input-output mapping of the neuronal dynamics with independent
variables $\nu$ and $CV$. Therefore, under the conditions described above,
a mean-field theory for the dynamics of the mean and variability of the spiking
response can be formulated. \citet{Doiron+06} have also recently used our
renormalization technique of the input variance, as defined in eqs.
(\ref{eq:sigma_renor},\ref{eq:nu_out-CV_out}), to describe 
the transmission of the activity of non-leaky IF neuron
in feed-forward networks. As we have said above, \citep{Renart+07}
have addressed the problem of self-consistency in firing rate and $CV$ in
recurrent networks of LIF neurons.
Other works have also studied this problem using different approaches
to find self-consistent equations for the spiking variability of the network
\citep{Lerchner+06}. 

However, further extensions of
our mean-field theory \citep{Ren+01,Mor+02,Renart+07}
are required to consider in a self-consistent way 
the second order statistics of the neuronal activity in spiking recurrent
networks. A first step has been made in \citep{Mor+06}, where 
the auto- and cross-correlation functions of the output response of a pair of spiking 
neurons receiving independent as well as common sources of noise have been 
analytically determined \footnote{For different approximations of this computation see
\citep{Lindner+05,Masuda06}}. 
The self-consistent treatment of spike cross-correlation functions (i.e., the
input and output cross-correlation functions should also match each other) to
describe more realistic recurrent neuronal networks seems to be an unavoidable
step to understand how neurons' interactions give rise to network behaviors.
The problem can be
formally stated as follows: find the set of mean-field equations mapping 
the input values of the relevant dynamical
variables of the network (firing rate, $F_N$ and $\rho$) to their output 
values

\bea
 \nu_{out} &=& f_{\nu}( \nu_{in}, F_{N,in}, \rho_{in}  )
 \nonumber
\\
 F_{N,out} &=& f_{F_N} ( \nu_{in}, F_{N,in}, \rho_{in}  )
 \nonumber
\\
 \rho_{out} &=& f_{\rho} ( \nu_{in}, F_{N,in}, \rho_{in}  )
 \nonumber
 \; .
\eea

\noindent
This set of equations are now available at least for a LIF neuron receiving
colored noise \citep{Mor+06}.

In this work we have considered decaying (exponential)
correlations, while in cortex, damped oscillatory cross-correlograms 
are also observed (see e.g. \citep{Vaadia+95,Rie+97,Fries+97}).  
This problem could be addressed by introducing 
a stochastic current which obeys a second-order
equation driven by white noise: the well-known damped oscillator. 
A current generated in this way 
can have a cross-correlogram with exponentially decaying oscillations, with frequency
and damping value controlled by the parameters of the equation.
Although relevant, we do
not study this problem here, since the new system would involve solving 
a more complicated FPE having now three independent variables. 

Here we have not studied neuron models with conductance-based synapses either. 
However, an analogous expression for the firing rate at long $\tau_c$ can be obtained if
the noise enters multiplicatively, instead of additively (although we do not
present the derivation here, the FPE for 
neuron models with conductance-based synapses can be
solved using the techniques in Appendix \ref{apendix_long_second}).
Qualitatively, the effect of the correlation magnitude and correlation timescale in
conductance-based models is not different from their effects in current-based
models. Note, however, that in the first case
correlations are strongly effective only when $\tau_c$ is shorter than the
effective membrane time constant of the neuron, which now depends on the total
conductance (see e.g. \citep{Mor+05}).

We have modelled input spike trains as delta functions (point-processes)
without any further temporal synaptic filtering. This means that the
cross-correlation function of the total input current displays a delta
function at zero time lag, as shown in eq. (\ref{eq:current_correlation}).
When the input spike trains are filtered by synapses with a finite synaptic time
constant $\tau_s$, they generate a train of exponential-like current waveforms
into the neuron. The delta function in the correlation function, 
eq. (\ref{eq:current_correlation}), becomes then an
exponential with the same time constant as that of the synaptic filter, $\tau_s$ (see
e.g. \citep{Bru+98b,Mor+04}). At the same time, the exponential term in the
correlation function results after filtering in two additional exponentials,
with time constants $\tau_c$ and $\tau_s$, respectively. 
Then, the result of (linearly) filtering correlated
input spike trains is an input current whose correlation function has two kind of
exponentials, each with a different time constant ($\tau_c$ and $\tau_s$). 
Particular cases of this interesting problem (e.g. when
the two timescales are disparate) could be addressed analytically by using
the techniques developed to study simultaneous 
fast and slow synaptic filtering \citep{Mor+04}. 

Two differences are expected when synaptic filters are present in the model.
First, synapses filter out fluctuations in the input whose timescale is shorter
than $\tau_s$ and convert them into fluctuations with timescale $\tau_s$. Fluctuations
that are slower than $\tau_s$ will pass the synapses. 
Therefore, fast fluctuations produced by precise 
input synchronization (i.e., short $\tau_c \le \tau_m$)
will not be seen by the neuron: effectively the sharp synchronization
of timescale $\tau_c$ is converted into a coarser synchronization with timescale $\tau_s$. 
Then, we expect that for $\tau_c < \tau_s$ the firing
rate will depend very little on $\tau_c$ in that range. 
However, when $\tau_c > \tau_s$ the
rate vs. $\tau_c$ curve will decay fast until $\tau_c$ crosses $\tau_m$,
after which the effect of input correlation on the rate will be small, similarly
to Fig. (\ref{fig:grafica1}).
Second, filters introduce a delay in the transient firing response to sudden
increases of input synchrony. 
We have run simulations with fast filters ($\tau_s \leq 5ms$) and found that
the response was still fast and was delayed by the time
constant of the synapses. 

In future work it would be desirable to develop a complete 
theory that describes the firing statistics of integrate-and-fire neurons with 
conductance-based synapses and finite synaptic timescales driven
by correlated spike trains. The effect of input correlations in this
more complex system could be evaluated by extending and combining the techniques
developed in this and the above quoted works.

\vspace{1cm}

\noindent
{\large\bf Acknowledgments }   

\vspace{0.5cm}
\noindent
Rub\'en Moreno-Bote thanks the Swartz Foundation for financial support.
Support was also provided by the Spanish Grant FIS 2006-09294

\vspace{1cm}


\appendix
 
{\huge\bf Appendices }


\section{Numerical procedures}
\label{numerical}

The equations for the voltage of the integrate-and-fire neuron 
and the correlated Gaussian noise are numerical solved using a simple 
Euler integration procedure, along with a Monte-Carlo method. This
procedure gives an excellent estimate of the
the output firing rate (time dependent or independent),
which can be compared to the theoretical predictions.
As an example, the dynamics
of the voltage of a LIF neuron in eq. (\ref{IF_equation}) with the current
$I(t)$ defined in eqs. (\ref{current},\ref{dynamic_z}) is integrated using
a small time step ($\delta t = 5 \; 10^{-4}ms \ll \tau_m$) as

\bea
&& V(t+\delta t) = V(t) - \frac{V(t)}{\tau_m} \delta t + I(t) \delta t\; ,
\\                                                             
&&I(t) = \mu +  \sigma_w \frac{\omega(t)}{  \sqrt{ \delta t } }+ 
					\sigma_w  \frac{\beta}{\sqrt{2\tau_c}}  z(t) \; ,
\\
&&z(t+\delta t) =  z(t) - \frac{z}{\tau_c}\delta t + 
					\sqrt{\frac{2}{\tau_c}} \omega(t) \sqrt{ \delta t} \; , 
\eea

\noindent
with the reset condition $V=H$ after a spike is generated (when $V \ge \Theta$). 
The initial value of the noise variable $z$ is that at the time of the previous spike,
i.e, $z$ is not reset after each spike.
The variable $\omega(t)$ is a random variable taken values $+1$ and $-1$ with equal
probability $1/2$ at each time step $\delta t$, and being drawn independently from
time step to time step. Therefore $\left < \omega(t)  \right> = 0 $, 
$\left < \omega^2(t)  \right> = 1 $ and $\left < \omega (t) \omega (t') \right> = 0 $,
where $t \neq t'$. This means that the quantity $\omega(t) /  \sqrt{ \delta t }$, which
appears in the expression for the current $I(t)$ above, is an approximation to the
delta function, since $ \left < \omega(t) /  \sqrt{ \delta t } \right> = 0$,
$ \left < \omega^2(t) /  \delta t  \right> = 1/\delta t$, and 
$ \left < \omega(t) \omega(t') /  \delta t  \right> = 0$. The procedure described
above is robust and converges to the true stationary process as $\delta t$ decreases.
The Monte-Carlo simulations were run using Fortran90 custom code. Special care
has to be taken in choosing an appropriate random generator for $\omega(t)$.


\section{Derivation of the FPEs}
\label{derivationFPE}

The FPE (\ref{FP-equation}) is here derived for the set of equations

\bea
 &&\dot{x}(t)=-\frac{x(t)}{\tau_m} + \sqrt{\frac{2}{\tau_m}} \eta(t) 
            + \frac{\beta}{\sqrt{ \tau_m \tau_c }} z(t)
\nonumber
\\
&&\dot{z}(t) =  - \frac{z}{\tau_c} + \sqrt{\frac{2}{\tau_c}} \eta(t) 
\; ,
\label{eq:x-z}
\eea

\noindent
corresponding to the first representation of the current.
The FPE (\ref{FP-equation2bis}) 
associated to the second representation of the current can be obtained
using the same rules described in this section. 
More formal derivations of similar FPEs can be found in \citep{Ric77,Ris89}. 

The system defined
by eqs. (\ref{eq:x-z}) is fully described by the probability density function
$P_{\beta}(x,z,t)$. This function expresses the probability density of having the
neuron in the state $(x,z)$ at time $t$. The FPE is an equation with precisely
describes the dynamics (i.e., time evolution) of such a density.
A first step toward the derivation of 
the FPE consists in discretizing the time in the dynamics, similarly as it has
been done in Appendix \ref{numerical}. This leads to 

\bea
 && x (t + \delta t)= x(t) -\frac{x(t)}{\tau_m} \delta t 
			+ \sqrt{\frac{2}{\tau_m}} \omega(t) \sqrt{ \delta t} 
            + \frac{\beta}{\sqrt{ \tau_m \tau_c }} z(t) \delta t
\nonumber
\\
&&z(t +  \delta t) = z(t) - \frac{z}{\tau_c} \delta t 
			+ \sqrt{\frac{2}{\tau_c}} \omega(t) \sqrt{ \delta t} 
\; ,
\label{eq:x-z-discre}
\eea

\noindent
where $\delta t$ represents an infinitesimal time increment, and $\omega(t)$ is
a random variable taken values $+1$ and $-1$ with probability $p(w= \pm 1)=1/2$ 
and drawn independently at
every infinitesimal time step. The terms in eqs. (\ref{eq:x-z-discre}) proportional
to $\sqrt{ \delta t} $ are approximations to the delta functions in eqs. (\ref{eq:x-z})
integrated during the infinitesimal time increment.

To determine the FPE associated to eqs. (\ref{eq:x-z}), one has to
relate the density at time $t+\delta t$, $P_{\beta}(x,z,t+\delta t)$, 
with the density at a previous time $t$, $P_{\beta}(x',z',t)$.
First, we realize that the probability that we find a neuron in an infinitesimal square 
$\delta x' \delta z'$ around state $(x',z')$ at time $t$ has
probability $P_{\beta}(x',z',t) \delta x' \delta z'$. 
Second, the state square centered at $(x',z')$ with surface $\delta x' \delta z'$
will be projected at the successive time $t+\delta t$ into another square
centered at $(x,z)$ with surface $\delta x \delta z$ close to the previous one, 
obeying the rules defined in eqs. (\ref{eq:x-z-discre}). Therefore, by
conservation of the probability, we have that

\be
  P_{\beta}(x,z,t + \delta t) \; \delta x \delta z 
     = \sum_{w= \pm 1} p(w) \;  
         P_{\beta}(x'(w),z'(w),t) \; \delta x' \delta z'   
         \; ,
  \label{eq:P_equal_dis}
\ee

\noindent 
where 

\bea
   && x'(w) = x + \frac{x}{\tau_m} \delta t 
			- \sqrt{\frac{2}{\tau_m}} \omega \sqrt{ \delta t} 
            - \frac{\beta}{\sqrt{ \tau_m \tau_c }} z \delta t
  \nonumber
  \\
 && z'(w) = z + \frac{z}{\tau_c} \delta t 
			- \sqrt{\frac{2}{\tau_c}} \omega \sqrt{ \delta t} 
	\; .
  \nonumber
\eea

\noindent
Notice that the states $(x'(w),z'(w))$ ($w= \pm 1$) defined above
are the only ones from where one can arrive to the state $(x,z)$ after
an infinitesimal amount of time $\delta t$. In addition, the box around
state $(x',z')$ is compressed to the box around the final state $(x,z)$
by a factor $\delta x \delta y = 
(1 - \delta t / \tau_m) (1 - \delta t / \tau_c) \delta x' \delta y' $, given
by the decaying term in eqs. (\ref{eq:x-z}).

After expanding the densities in eq. (\ref{eq:P_equal_dis})
in powers of $\sqrt{  \delta t }$, we find that all terms which are
order $ \sqrt{ \delta t } $ are equal to zero (since $\left< \omega  \right> = 0 $),
while the terms order $\delta t$ do not vanish 
(either they do not depend on $\omega$, or they are proportional to $\omega^2$,
and therefore $\left< \omega^2  \right> = 1 $). After
equaling the terms at O($\delta t$), one obtains the FPE 
\begin{equation}
  \tau_m \frac{\partial}{\partial t} P_{\beta}(x,z,t) =  
   [L_x  + \frac{L_z}{k^2} 
           + \frac{2}{k} \frac{\partial}{\partial x}(
                \frac{\partial}{\partial z} -\frac{\beta z}{2})] 
          P_{\beta}(x,z,t)   
          \; .
 \nonumber
\end{equation}

In the time-independent case,
$\frac{\partial}{\partial t} P_{\beta}(x,z,t) = 0$.
However, to establish a stationary 
probability density function which does not depend on time, 
the probability density flux escaping at threshold 
(probability density flux in the direction
of the variable $x$ calculated at threshold) 
should be reinjected into the reset voltage. This enforces conservation of
the total probability, that is, $ \int \int dx dz \rho(x,z,t)=1$ at all times,  
and leads to the
self-consistent stationary FPE (\ref{FP-equation}).


\section{Long $\tau_c$ expansion using the first representation}
\label{apendix_long}

Here we detail the main steps for calculating the firing rate in eq.
(\ref{nu_out_big_tauc}). Introducing the expansions
(\ref{eq:P_beta_expansion}, \ref{eq:J_beta_expansion}) in eq.
(\ref{FP-equation}) we obtain

\be
 L_x h_n + L_z h_{n-2} + 2 \frac{\partial}{\partial x}(
                \frac{\partial}{\partial z} -\frac{\beta z}{2}) h_{n-1}
             + \tau_m \delta(x-\sqrt{2}\hat H) J_{\beta,n-1}(z) = 0 
 \label{FP123}
\ee

\noindent
($h_n \equiv 0$ for $n<0$). The solution to these equations is obtained
order by order in such a manner that the conditions (\ref{condition1}
- \ref{eq:escape_probability}) are satisfied. After solving them up to order
$k^2$ using the conditions (\ref{condition2},
\ref{eq:escape_probability}) and the fact that the $h_n$'s have to be
normalizable, we obtain that

\bea
&&h_0(x,z) = k_0(x) J_{\beta,0}(z)
 \; , 
\nonumber
\\
&&h_1(x,z) = k_0(x) J_{\beta,1}(z) 
         +  k_1(x) (2 \frac{\partial}{\partial z} - \beta z)  J_{\beta,0}(z)
 \; , 
\nonumber
\\
&&h_2(x,z) = k_0(x) J_{\beta,2}(z) + k_1(x)  
          (2 \frac{\partial}{\partial z} - \beta z)   J_{\beta,1}(z)
 \nonumber
\\
 && \;\;\;\;\;\;\;       
   + k_2(x) (2 \frac{\partial}{\partial z} - \beta z)^2  J_{\beta,0}(z)
    \; , 
\label{eq:h012}
\eea

\noindent
where the functions $k_i$ are 

\bea
&&k_0(x) = \tau_m e^{-\frac{x^2}{2}}
         \int_{x}^{\sqrt{2} \hat \Theta} 
         dy e^{\frac{y^2}{2}} H(y-\sqrt{2}\hat H)
 \; , 
\nonumber
\\
&&k_1(x) = e^{-\frac{x^2}{2}}
         \int_{x}^{\sqrt{2} \hat \Theta} dy e^{\frac{y^2}{2}}  k_0(y)
 \; , 
\nonumber
\\ 
&&k_2(x) = e^{-\frac{x^2}{2}}
         \int_{x}^{\sqrt{2} \hat \Theta} dy e^{\frac{y^2}{2}}  k_1(y)
 \; . 
\nonumber
\eea

\noindent
The coefficients $J_{i,\beta}(z)$ in eq. (\ref{eq:h012})
have still to be calculated. This
is done by integrating first the $h_n$'s over $x$ from $-\infty$ to $\sqrt{2
}\hat \Theta$ and using the condition (\ref{eq:Gauss-z-y}). After
using the condition (\ref{condition3}), we find

\bea
&&J_0(z)=\nu_0 Z_0(z)
 \; , 
\nonumber
\\
&&J_1(z)=\frac{ (2+\beta) \nu_0 
\int_{-\infty}^{\sqrt{2}\hat \Theta} dx k_1(x)}
{\int_{-\infty}^{\sqrt{2}\hat \Theta} dx k_0(x)} \; z\; Z_0(z)
 \; , 
\nonumber
\\
&&J_2(z) = \left[ \frac{\alpha}{\tau_m} C  +
 \frac{\alpha C}{\beta^2 \tau_m (1-\nu_0 \tau_{ref})}
        (z^2-1) \right] Z_0(z)
 \; , 
\nonumber
\\
&&C = \tau_m \nu_0^2 
\left[ \frac{(\int_{-\infty}^{\sqrt{2}\hat \Theta} dx k_1(x))^2}
{\int_{-\infty}^{\sqrt{2}\hat \Theta} dx k_0(x)} -
\int_{-\infty}^{\sqrt{2}\hat \Theta} dx k_2(x)  
 \right] 
 \; ,
\label{eq:j0-j2-c}
\eea

\noindent
where $Z_0(z)=e^{-z^2/2}/\sqrt{2 \pi}$. Finally, integrating again
$J_{i,\beta}(z)$ over $z$ gives the contributions to the output firing rate
in eq. (\ref{nu_out_big_tauc}). In the next section we calculate the
integrals appearing in the parameter $C$ in eq. (\ref{eq:j0-j2-c}).

\subsection{Integrals}

Here we only present some intermediate steps and the final results for
the integrals appearing in $C$, eq. (\ref{eq:j0-j2-c}). 
The last two integrals can be expressed
in terms of the function $R(t)= \sqrt{\frac{\pi}{2}} e^{t^2}(1 +
\rm{erf}(t))= e^{t^2}\int_{-\infty}^{\sqrt{2} t} ds \; e^{-s^2/2}$ as

\bea
 1. &&\int_{-\infty}^{\sqrt{2}\hat \Theta} dx k_0(x) = 
\nonumber
\\
 && \;\; \tau_m \int_{-\infty}^{\sqrt{2}\hat \Theta} dx e^{-\frac{x^2}{2}}
 \int_{x}^{\sqrt{2} \hat \Theta} dy e^{\frac{y^2}{2}} H(y-\sqrt{2}\hat H) =
 \frac{1- \nu_0 \tau_m}{ \nu_0} 
 \; .
\nonumber
\\
2. && \int_{-\infty}^{\sqrt{2}\hat \Theta} dx k_1(x) = 
 \int_{-\infty}^{\sqrt{2}\hat \Theta} dx  e^{-\frac{x^2}{2}}
 \int_{x}^{\sqrt{2} \hat \Theta} dy e^{\frac{y^2}{2}}  k_0(x) =
\nonumber
\\
&& \;\; \tau_m \int_{\sqrt{2}\hat H }^{\sqrt{2}\hat \Theta}
 dy e^{\frac{y^2}{2}} 
 \int_{-\infty}^{y} dx  e^{-\frac{x^2}{2}} (y-x) =
 \tau_m (R(\hat \Theta)-R(\hat H))
 \; . 
\nonumber
\\
3. && \int_{-\infty}^{\sqrt{2}\hat \Theta} dx k_2(x) = 
 \int_{\sqrt{2}\hat H}^{\sqrt{2}\hat \Theta}  
 \frac{\tau_m}{2} dy e^{\frac{y^2}{2}} 
 \int_{-\infty}^{y} dx  e^{-\frac{x^2}{2}} (y-x)^2 =
\nonumber
\\
 && \;\; \tau_m ( \frac{\hat \Theta}{\sqrt{2}}R(\hat \Theta) - 
 \frac{\hat H}{\sqrt{2}}R(\hat H) )
 \; . 
\nonumber
\eea


\section{Long $\tau_c$ expansion using the second representation}
\label{apendix_long_second}

In this section we derive the output firing rate formula
(\ref{nu_out_big_tauc-gamma}) using the FPE (\ref{FP-equation2bis}).
Here, we take the ratio $\gamma \equiv \sqrt{\alpha}/k$ to be a
parameter independent of $k$, that is, it is fixed. This will allow us
to study the case of large $\alpha$ in the long $\tau_c$ limit.
From the FPE (\ref{FP-equation2bis},\ref{FP-equation-gamma}) 
we develop a systematic expansion of the probability
distribution $P_{\alpha}(x,y)$ and the escape probability density flux
$J_{\alpha}(y)$ in powers of $k^{-2}$ (see the expansion in
(\ref{eq:P-J})), in which $\gamma$ is considered a fixed parameter 
independent of $k$. Inserting the expansion in eq. (\ref{eq:P-J}) into the FPE
(\ref{FP-equation-gamma}) produces

\be
[L_x-\gamma y \frac{\partial}{\partial x} ]r_n + L_y r_{n-1} 
             + \tau_m \delta(x-\sqrt{2}\hat H) J_{\alpha,n}(y) = 0  \;\;,
 \label{eq:FPE-expansion-gamma} 
\ee

\noindent
where $r_n \equiv 0$ if $n < 0$.  For simplicity, the set of conditions
(\ref{condition1} - \ref{eq:Gauss-z-y}) is used here when
$\tau_{ref}=0$. Solving the zero-th order in eq.
(\ref{eq:FPE-expansion-gamma}) with conditions (\ref{condition2},
\ref{eq:escape_probability}) gives

\be
 r_{\alpha,0}(x,y)=\tau_m J_{\alpha,0}(y) \; e^{-\frac{(x-\gamma y)^2}{2}}
 \int_{x}^{\sqrt{2} \hat \Theta} 
   du \; e^{\frac{(u-\gamma y)^2}{2}} H(u-\sqrt{2}\hat H)  \; ,
\nonumber
\ee

\noindent
where the escape probability density flux $J_{\alpha,0}(y)$ has yet to be
determined.  This is done by using the condition (\ref{eq:Gauss-z-y})
with $\tau_{ref}=0$ at zero-th order, to obtain

\be
 J_{\alpha,0}(y) = \frac{1}{\sqrt{2 \pi} \tau_m} e^{-\frac{y^2}{2}}
    \left[ \int_{\sqrt{2} \hat H -\gamma y}^{\sqrt{2} \hat \Theta -\gamma y}  
       du e^{\frac{u^2}{2}} 
       \int_{-\infty}^{u} dv e^{-\frac{v^2}{2}} 
    \right]^{-1} 
    \; .
 \label{eq:J-zero-order}
\ee

\noindent
Repeating the same steps as above, the $n-th$ ($n > 0$) order
escape probability density flux is found to be

\bea
&& J_{\alpha,n+1}(y) = \left[ \int_{\sqrt{2} \hat H -\gamma y}^{\sqrt{2} 
       \hat \Theta -\gamma y}  
       du e^{\frac{u^2}{2}} 
       \int_{-\infty}^{u} dv e^{-\frac{v^2}{2}}
      \right]^{-1}   
\nonumber
 \\
&&  \;\;\;\;\;\;     \int_{-\infty}^{\sqrt{2} \hat \Theta }  
       dx e^{\frac{-(x-\gamma y)^2}{2}} 
       \int_{x}^{\sqrt{2} \hat \Theta} dv e^{\frac{(v-\gamma y)^2}{2}}  
       L_y  \int_{-\infty}^{u} dv r_{n} (v,y)
      \; .
 \label{eq:J-order-general}
\eea

\noindent
and the density $r_n$ is computed as

\bea
&& r_n(x,z) = e^{\frac{-(x-\gamma y)^2}{2}} 
       \int_{x}^{\sqrt{2} \hat \Theta} dv e^{\frac{(v-\gamma y)^2}{2}}  
       L_y  \int_{-\infty}^{u} dv r_{n-1} (v,y)
 \nonumber
\\
&&  \;\;\;\;\;\; + \tau_m J_{\alpha,n}(y) \; e^{-\frac{(x-\gamma y)^2}{2}}
 \int_{x}^{\sqrt{2} \hat \Theta} 
   du \; e^{\frac{(u-\gamma y)^2}{2}} H(u-\sqrt{2}\hat H)  \; .
\nonumber
\eea

The zero-th order rate is obtained by integrating over $y$
the zero-th order escape probability density flux in eq. (\ref{eq:J-zero-order})
This gives the firing rate in eq. (\ref{nu_out_big_tauc-gamma}).
For fixed $\alpha$, the parameter $\gamma$
decreases as $\tau_c$ grows. In this
limit, we could expand the zero-th order firing rate in powers of
$\gamma$. The firing rate obtained from
this expansion has a dominant order $k^{-2}$ ($O(\gamma^{2})$).
However, other contributions to the total firing rate at order
$k^{-2}$ could also come from the non zero-th order firing rate
from the expansion (\ref{eq:P-J}). In particular, the first order ($n=1$) rate
in the expansion (\ref{eq:P-J}) is order $k^{-2}$.  However, it is
possible to see that an expansion in powers of $\gamma$ in the term
with $n=1$ in eq.  (\ref{eq:J-order-general}) also leads to an extra
dominant order $k^{-2}$, that multiplied by $k^{-2}$ yields finally
a correction to the firing rate bigger than $O(k^{-2})$. This finally
proves that the leading correction to the firing rate for fixed $\alpha$ when $\tau_c$
approaches infinity is order $k^{-2}$ and it is given by the
expansion of the zero-th order rate (\ref{nu_out_big_tauc-gamma}).
Naturally, this expansion matches the output firing rate
formula (\ref{nu_out_big_tauc}) for positive correlation magnitudes.


\section{Short $\tau_c$ expansion using the second representation}
\label{appendix:short}

\subsection{The Free Solution}

\noindent
We introduce an expansion of the form (\ref{P_alpha_expansion},
\ref{nu_alpha_expansion}) into the FPE (\ref{FP-equation2bis}) and find
the set of equations:

\bea
&&L_y f_0  = 0 
\; ,
\label{eq1}
\\
&&L_y f_1 = \sqrt{\alpha} y f_0 
\; , 
\label{eq2}
\\
&&L_y f_2 =  - L_x f_0 + \sqrt{\alpha} y f_1
 - \tau_m \delta(x- \sqrt{2} \hat H)   \nu_{eff}Z_0(y)
 \; , 
\label{eq3}
\\
&&L_y f_3 =  - L_x f_1 + \sqrt{\alpha} y f_2 
 - \tau_m \delta(x- \sqrt{2} \hat H)    \nu_{1} Z_0(y)
 \; , 
\label{eq4}
\eea 

\noindent
where $Z_0(y)=e^{-y^2/2}/\sqrt{2 \pi}$.
After solving eq. (\ref{eq1}) we find that the only normalizable solution is 

\be
f_0(x,y)=g_0(x) Z_0(y)
 \; , 
\label{eq:f_0_appen}
\ee

\noindent
where $g_0$ has yet to be determined. The equation at order $k$ gives the
expression for $f_1$

\be
 f_1(x,y)=[g_1(x)- \sqrt{\alpha} y  
   \frac{\partial}{\partial x} g_0(x)] Z_0(y)
 \; .
\label{eq:f_1_appen}
\ee

\noindent
Again, $g_1$ is unknown. The equation at second order satisfies

\bea
&& L_y f_2(x,y) =
      -\alpha y g_1(x) Z_0(y)
 \nonumber
\\
   && \;\;\;\;\;\;  
   -[L_x g_0(x) + \alpha \frac{\partial^2 }{\partial x^2 } g_0(x) 
               -\tau_m \delta(x-\sqrt{2}\hat H) \nu_{eff}] Z_0(y)
 \; .
\label{Lzf2}
\eea

\noindent
Using that the integral $\int dy L_y f_2(x,y)$ has to equal zero in order for
$f_2$ to be integrable, we can integrate eq. (\ref{Lzf2}) over $y$ and
obtain the condition

\be
[\frac{\partial }{\partial x} x +
(1+\alpha) \frac{\partial^{2} }{\partial^{2} x}] g_0(x) +
\tau_m \nu_{eff} \delta(x- \sqrt{2} \hat H) 
=0
\; .
\label{equation-g0}
\ee

\noindent
This equation is the same as that obtained when solving the FPE for a
LIF neuron driven by white noise input \citep{Ric77}, but where the
variance of the noise has been renormalized by a factor $1+\alpha$.
This equation is solved exactly for all $\alpha$ using the condition
(\ref{condition2}):

\be
 g_0(x)=\frac{\tau_m \nu_{eff}}{1+\alpha} e^{-\frac{x^2}{2(1+\alpha)}}
         \int_{x}^{\sqrt{2} \hat \Theta} dy e^{\frac{y^2}{2(1+\alpha)}} 
         H(y-\sqrt{2}\hat H)
         \;\;
         .
 \nonumber
\ee

\noindent
The firing rate $\nu_{eff}$ (the zero-th order in the expansion in powers of $k$)
is obtained by applying the condition (\ref{condition1}) to $f_0$
in eq. (\ref{eq:f_0_appen}).

Similarly, while solving eq. (\ref{eq4}) a condition over $g_1$ is obtained,
from where $g_1$ is determined, except for an unknown constant $D$:

\be
 g_1(x)= D e^{-\frac{x^2}{2(1+\alpha)}}
         +
          \frac{\tau_m \nu_1}{1+\alpha} e^{-\frac{x^2}{2(1+\alpha)}}
         \int_{x}^{\sqrt{2} \hat \Theta} dy e^{\frac{y^2}{2(1+\alpha)}} 
         H(y-\sqrt{2}\hat H)
  \; .
\label{equation-g1}
\ee

\noindent
The constant $D$ is needed to match the boundary condition at
threshold (\ref{condition2}).  

Now it is crucial to realize that the first order solution
$f_1$ does not satisfy the boundary condition at threshold
(\ref{condition2}) for any value of $D$. Thus, we have to add a {\it
  boundary solution} $f_1^b$ so that the total solution
(\ref{P_alpha_total}) satisfies it up to order $k$. This boundary
solution, found in the next section, serves to fix the value for $D$
as

\be
 D=\alpha \; \nu_{eff} \; \tau_m \; e^{\frac{\hat \Theta^2}{(1+\alpha)}}.
\label{value_d}
\ee

\noindent
Using the normalization condition (\ref{condition1}) on the term
order $k$ in the expansion of $P_{\alpha}(x,y)$, eq.
(\ref{P_alpha_total}) leads to the firing rate at order $k$ 
in eq. (\ref{nu_out_small_tauc}) \footnote{Notice below that
  $\int \int u(r,z)= O(k)$, and for this reason we can neglect its contribution
  to the rate at order $k$.}.
In that equation we have approximated $\nu_{eff}$ by $\nu_0$ and also all $\alpha$
appearing in eq. (\ref{equation-g1}) have been made equal to zero. These
two approximations are justified because expanding $\nu_{eff}$ and eq.
(\ref{equation-g1}) in powers of $\alpha$ gives corrections to the
firing rate at order $k$ that are higher than $O(\alpha)$.

\subsection{The Boundary Solution}
\label{subsec:boundary}

Here we find the boundary solution, $f_1^b$, for the FPE
(\ref{FP-equation2bis}) valid close to threshold and for small $k$. The
FPE in this limit takes the form

\be
\left[ \frac{\partial^2}{\partial r^2} - \sqrt{\alpha} y \frac{\partial}{\partial r}
+\frac{\partial^2}{\partial y^2} - y \frac{\partial}{\partial y}
+ O(k,k^2) \right] u(r,y) =  0
\; .
\label{FPr_appen}
\ee

\noindent
We have replaced $f_1^b(x,y)= u(r,y) Z_0(y)$ and we have made the linear
transformation $r=(x-\sqrt{2} \hat \Theta)/k$.  A complete basis for this 
linear differential operator is not known, but if
$\sqrt{\alpha}=0$ a complete basis for an integrable function of $r
\in [-\infty,0], y \in [-\infty,\infty]$ is given by the set of
functions $e^{\sqrt{n} r} H_n(y/\sqrt{2})$ for all $n>0$, where $H_n$
are the Hermite polynomials 
\footnote{The Hermite polynomials satisfy
  the equation

\be
 \left(  \frac{\partial^2}{\partial y^2} - 
 y \frac{\partial}{\partial y} \right) H_n \left( \frac{y}{\sqrt{2}} \right)
 = 
 -n \; H_n \left( \frac{y}{\sqrt{2}} \right)
 \;.
 \nonumber
\ee

\noindent
The first three polynomials $H_0(y)=1$,
$H_1(y)=2y$ and $H_2(y)=4y^2-2$ are used in our calculations.

}.  We insert into eq.  (\ref{FPr_appen}) a solution $u$ of the form $u=u_0
+\sqrt{\alpha} u_1 + \alpha u_2 + O(\alpha^{3/2})$ to obtain

\be
\left[ \frac{\partial^2}{\partial r^2} 
+\frac{\partial^2}{\partial y^2} - y \frac{\partial}{\partial y}
\right] u_{i+1}(r,y) =  y \frac{\partial}{\partial r}  u_i(r,y)
 \; .
\label{FPr}
\ee

\noindent
The solution $f_1^b$ has to be added to the perturbative solution
$f_1$, eq (\ref{eq:f_1_appen}), to match the 
boundary condition (\ref{condition2}), that is

\be
D e^{-\frac{\hat \Theta^2}{1+\alpha}}-
\sqrt{\alpha} y \frac{\partial}{\partial x}g_0|_{x=\sqrt{2}\hat \Theta}+
u(0,y)
= 0
 \; .
\label{cond_D}
\ee

\noindent
Defining $d=D e^{-\frac{\hat \Theta^2}{1+\alpha}}$ and expanding it
in powers of $\sqrt{\alpha}$ as $d=d_0 + \sqrt{\alpha} d_1 + \alpha
d_2 + O(\alpha^{3/2})$, as well as the others terms in eq. (\ref{cond_D}), 
we obtain the set of conditions

\bea
&&d_0 + u_0(0,y) = 0
 \; , 
\nonumber
\\
&&d_1 + \nu_{eff} \tau_m y +  u_1(0,y) = 0
 \; , 
\nonumber
\\
&&d_2 + u_2(0,y) = 0
\; .
\nonumber
\eea

\noindent
Now we express each order $u_i$ as a linear combination of the
functions $e^{\sqrt{n} r} H_n(y/\sqrt{2})$ plus a particular solution
as $u_i(r,y)=\sum_{1}^{\infty} A_{n,i} e^{\sqrt{n} r} H_n(y/\sqrt{2})
+ u_{i,part}(r,y)$. We find

\bea
&&u_0=0  \;,\;\;\;  d_0=0
\nonumber
\\
&&u_1= -\nu_{eff} \tau_m y e^r  \;,\;\;\; d_1=0
\nonumber
\\
&&u_2= -\nu_{eff} \tau_m [y^2-1] e^{ \sqrt{2} r}  + \nu_{eff} \tau_m [y^2-2]e^r
\;,\;\;\;  d_2 = \nu_{eff} \tau_m
\; .
\nonumber
\eea

\noindent
With these solutions, we finally found the value of $D$ up to order
$\alpha$, eq. (\ref{value_d}).


\section{Short $\tau_c$ limit for a generic IF neuron.}
\label{appendix:short-generic}

In this section we extend the formalism described in Appendix
\ref{appendix:short} to calculate the firing rate of a generic IF
neuron receiving a Gaussian exponentially correlated input in the
short $\tau_c$ limit \citep{Mor+02a}. A generic IF neuron can be
defined by the leak function, $f(V)$, that determines how the voltage
behaves in absence of any input. In this model, the depolarization
membrane potential $V(t)$ evolves from the reset voltage $H$ according
to the stochastic equation

\begin{equation}
\dot{V}(t) =  -f(V) + I(t)  \; ,
\label{IF_equation-generic}
\end{equation}

\noindent
where $I(t)$ is the synaptic current with exponentially temporal
correlations as in eq. (\ref{eq:current_correlation}).  When the
Gaussian current is expressed using the second representation, as it
is defined in Section (\ref{subsec:pos_corr}), the FPE associated to
this model neuron is

\begin{equation}
\left[ \frac{\partial}{\partial V} (f(V)-\mu+ 
       \frac{\sigma_w^2}{2} \frac{\partial}{\partial V} )  
     + \frac{1}{\tau_c}  \frac{\partial }{\partial y} 
         (  y + \frac{\partial}{\partial y} ) 
 - \sqrt{\frac{2 \sigma_w^2 \alpha }{\tau_c}} 
   \frac{\partial}{\partial V} \right] P
       =  
       -  \delta(V-H) J(y)  \; .
\label{FP-equation-generic}
\end{equation}

\noindent
Using the same procedure as in Appendix \ref{appendix:short},
we find that the output firing rate of such a generic neuron is

\be
   \nu_{out}=\nu_{eff}+\nu_1 \sqrt{\tau_c}
\ee

\noindent
where

\begin{eqnarray}
&&  \nu_{eff}^{-1}  = \tau_{ref} +  \frac{2}{\sigma^2_{eff}}
 \int_{H}^{\Theta} du e^{ \frac{2}{\sigma^2_{eff}} 
         \int_{\Theta}^{u} dr (f(r) -\mu) }
        \int_{-\infty}^{u} dv e^{ -  \frac{2}{\sigma^2_{eff}} 
         \int_{\Theta}^{v} dr ( f(r) -\mu) }
\nonumber \\
&& \nu_{1}  =  - \frac{\sqrt{2} \alpha \nu_0^2 }{\sigma_w}   
        \int_{-\infty}^{\Theta} dv e^{ -  \frac{2}{\sigma_w^2} 
         \int_{\Theta}^{v} dr ( f(r) -\mu ) }
         \; ,
\label{rates_teoricos.eq}
\end{eqnarray}

\noindent 
which is valid whenever the above integrals are defined.  This general
formula, that has been previously found in our work \citep{Mor+02a},
shows that the $\sqrt{\tau_c}$ decay of the firing rate is {\em
  universal} for IF models with hard threshold.  Using this general
formula it is possible to obtain the firing rate in the short $\tau_c$
limit given by eq.  (\ref{nu_out_small_tauc}) for a LIF neuron.

Using a different procedure we have been able to calculate exactly the
firing rate of a non-leaky IF neuron ($f(V)=0$) with exponential
correlations without the need of the boundary solution to fit the
boundary condition at threshold. This formula is valid for {\em all}
$\tau_c$ and for small $\alpha$. We still require the condition
$\tau_c \ll \tau_{ref}$.  This exact formula, however, allows us to check
the technical procedure described above, and it naturally gives the
same result. This firing rate is expressed as

\begin{equation}
\nu_{out}=\nu_{eff} - 
 \frac{ \alpha \nu_0^2  [1-e^{(\gamma-\lambda) (\Theta-H)}]}
 {\mu (\gamma + \lambda)}
 + O(\alpha^2)
\label{rates_nonleaky_todotau.eq}
\end{equation}

\noindent
where $\gamma=\frac{\mu}{\sigma_w^2}$,
$\lambda=\sqrt{\gamma^2+\frac{2}{\sigma_w^2 \tau_c}}$ and $\nu_{eff}$ is
defined below, eq.(\ref{rates_nonleaky.eq}). An expansion of eq.
(\ref{rates_nonleaky_todotau.eq}) for small $\tau_c$ leads to the
same universal $\sqrt{\tau_c}$ decay law, and the coefficients
are identical to those produced by eqs. (\ref{rates_teoricos.eq}):

\begin{eqnarray}
\nu_{eff}^{-1}  &= &\tau_{ref} +  \frac{\Theta-H}{\mu}
\nonumber \\
\nu_{1}  &= & -\frac{\alpha \nu_0^2 \sigma_w}{\sqrt{2} \mu}
 \; .
\label{rates_nonleaky.eq}
\end{eqnarray}


\newpage

\begin{figure}
\begin{center}
\includegraphics[width=15cm,height=6cm,angle=0]{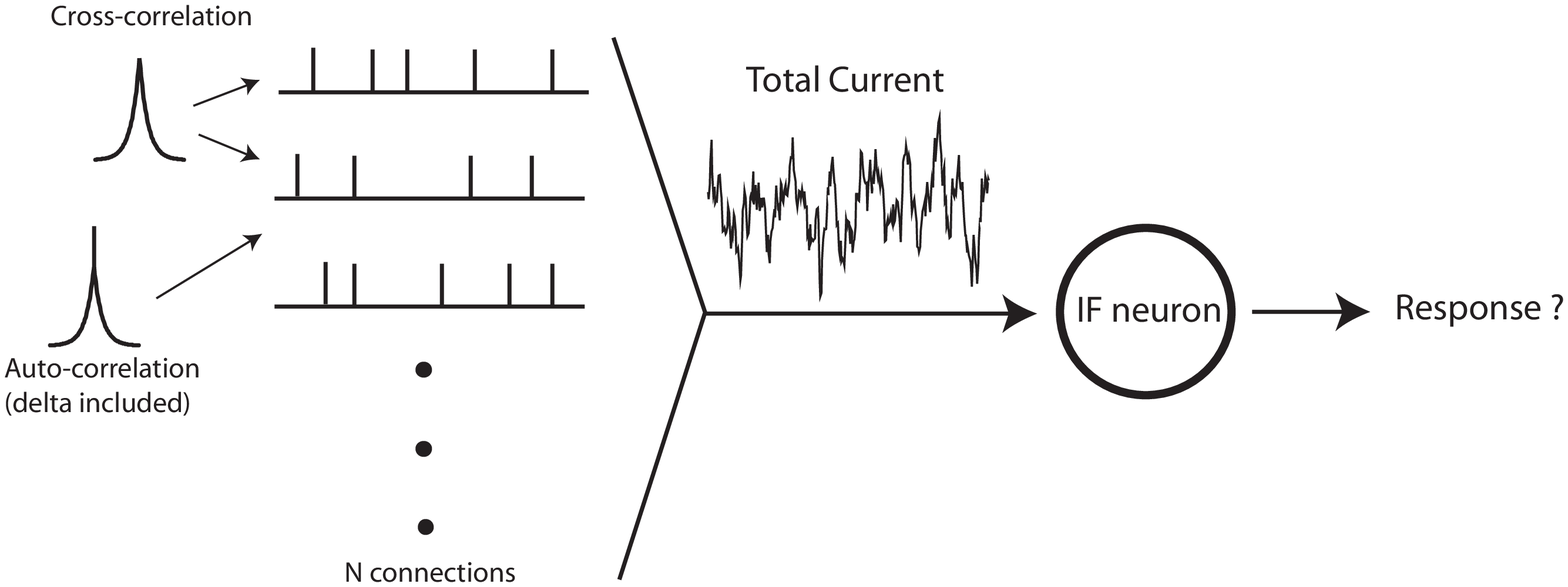}
\end{center}
\caption{\label{fig:scheme}
}
\end{figure}

\begin{figure}
\begin{center}
\includegraphics[width=10cm,height=14cm,angle=0]{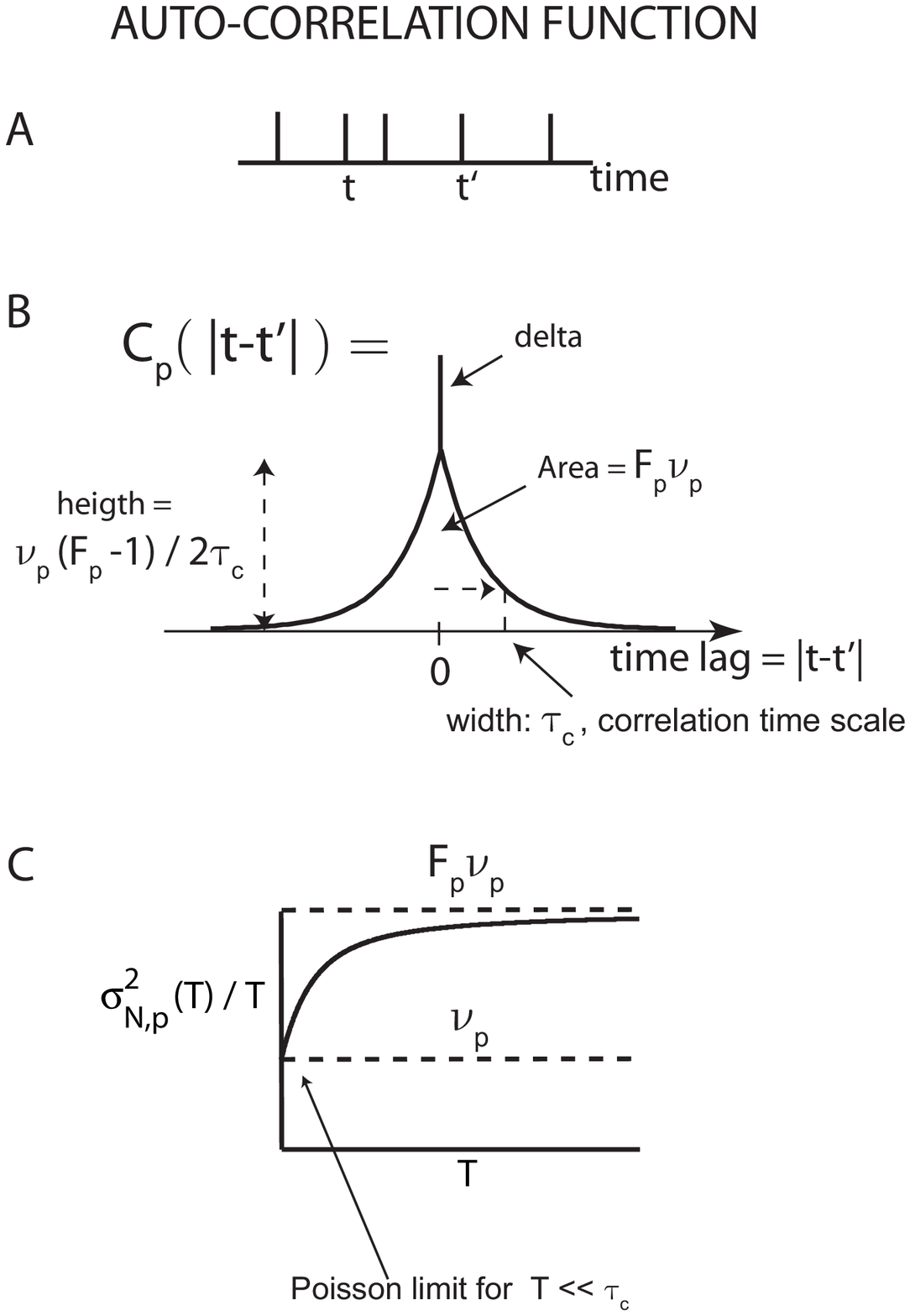}
\end{center}
\caption{\label{fig:FN}
}
\end{figure}

\begin{figure}
\begin{center}
\includegraphics[width=10cm,height=14cm,angle=0]{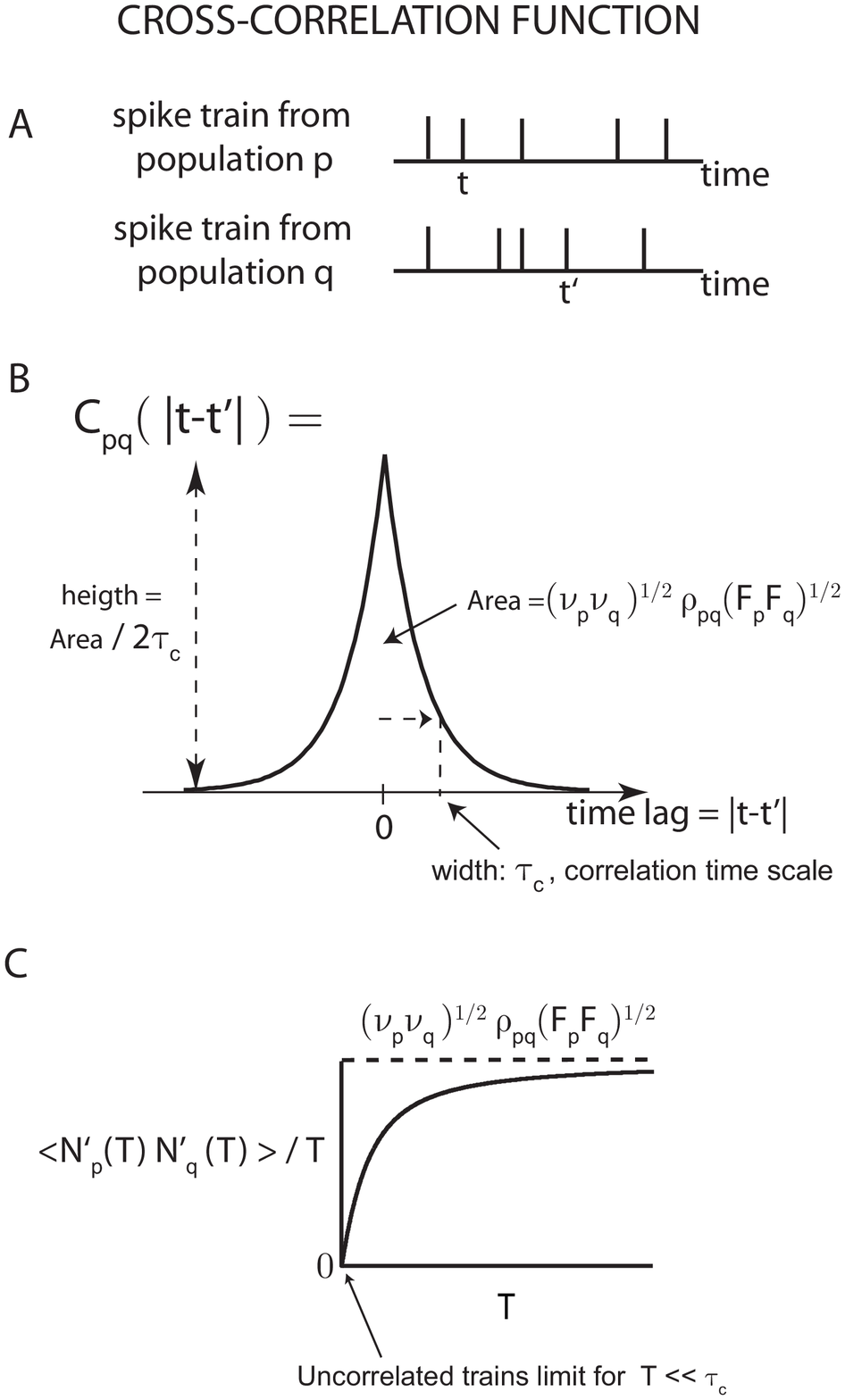}
\end{center}
\caption{\label{fig:rho}
}
\end{figure}

\begin{figure}
\begin{center}
\includegraphics[width=8cm,height=7.7cm,angle=0]{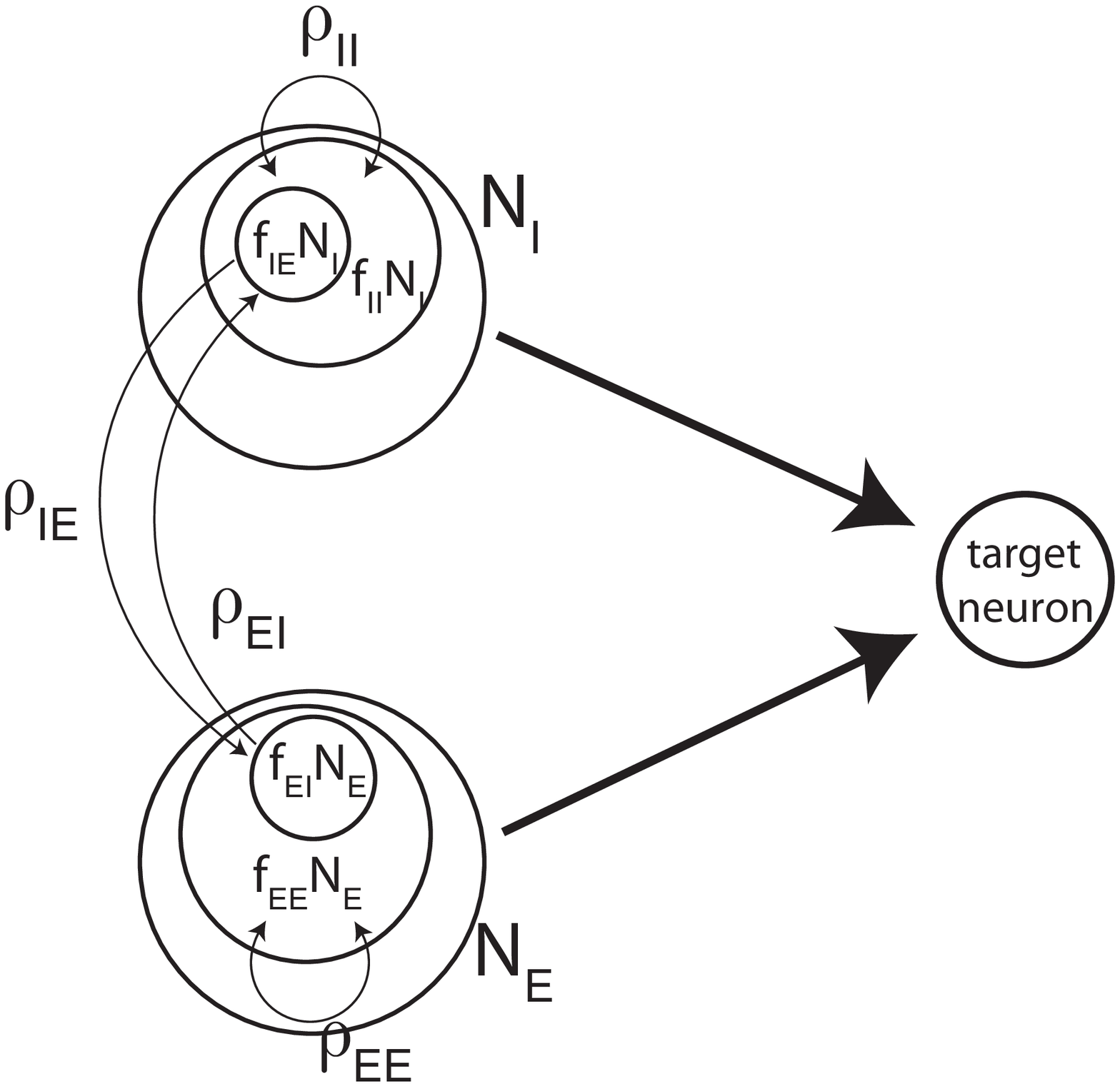}
\end{center}
\caption{\label{fig:archi}}
\end{figure}

\begin{figure}
\begin{center}
\includegraphics[width=10cm,height=7cm,angle=0]{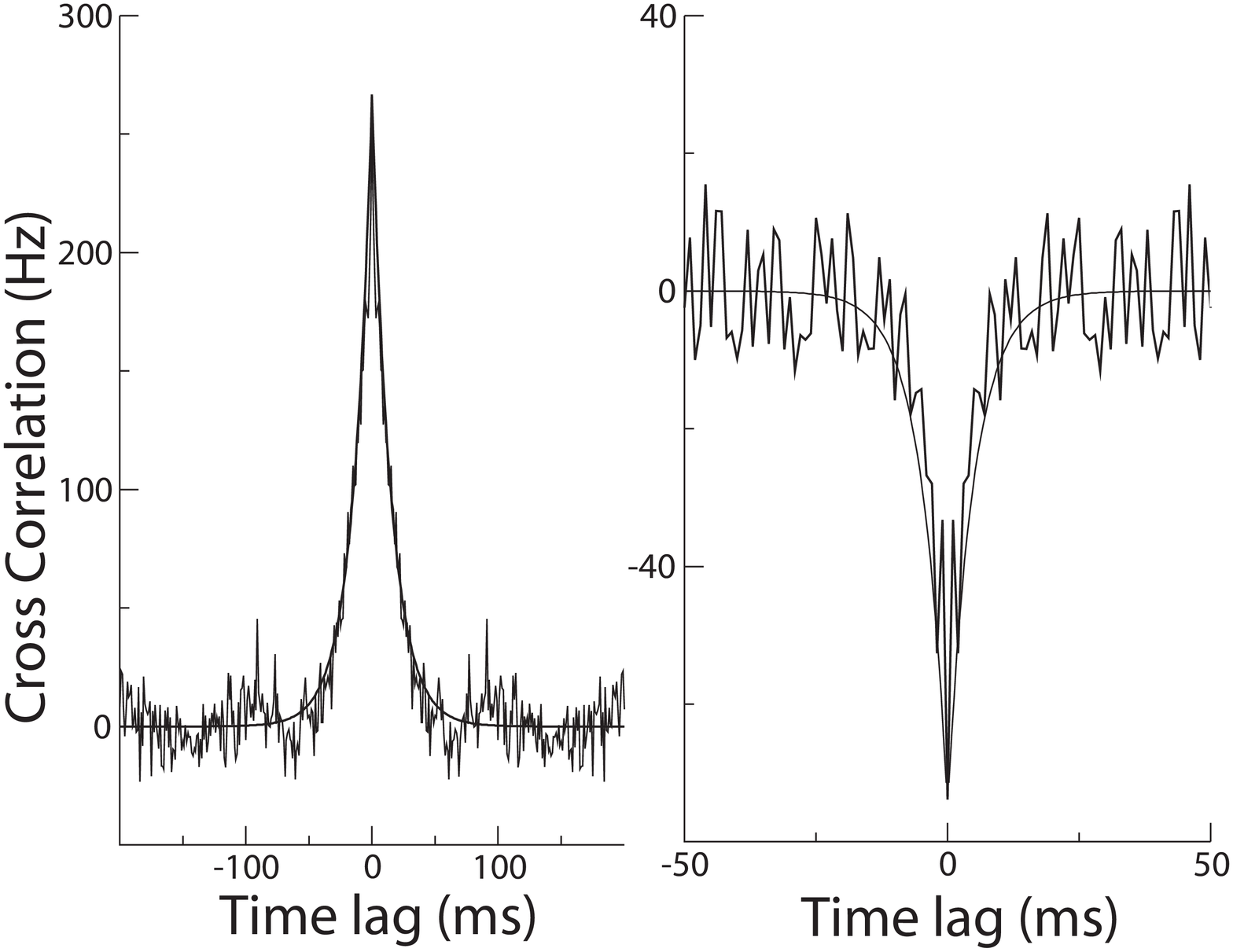}
\end{center}
\caption{\label{fig:correlogram}
}
\end{figure}

\begin{figure}
\begin{center}
\includegraphics[width=9cm,height=8cm,angle=0]{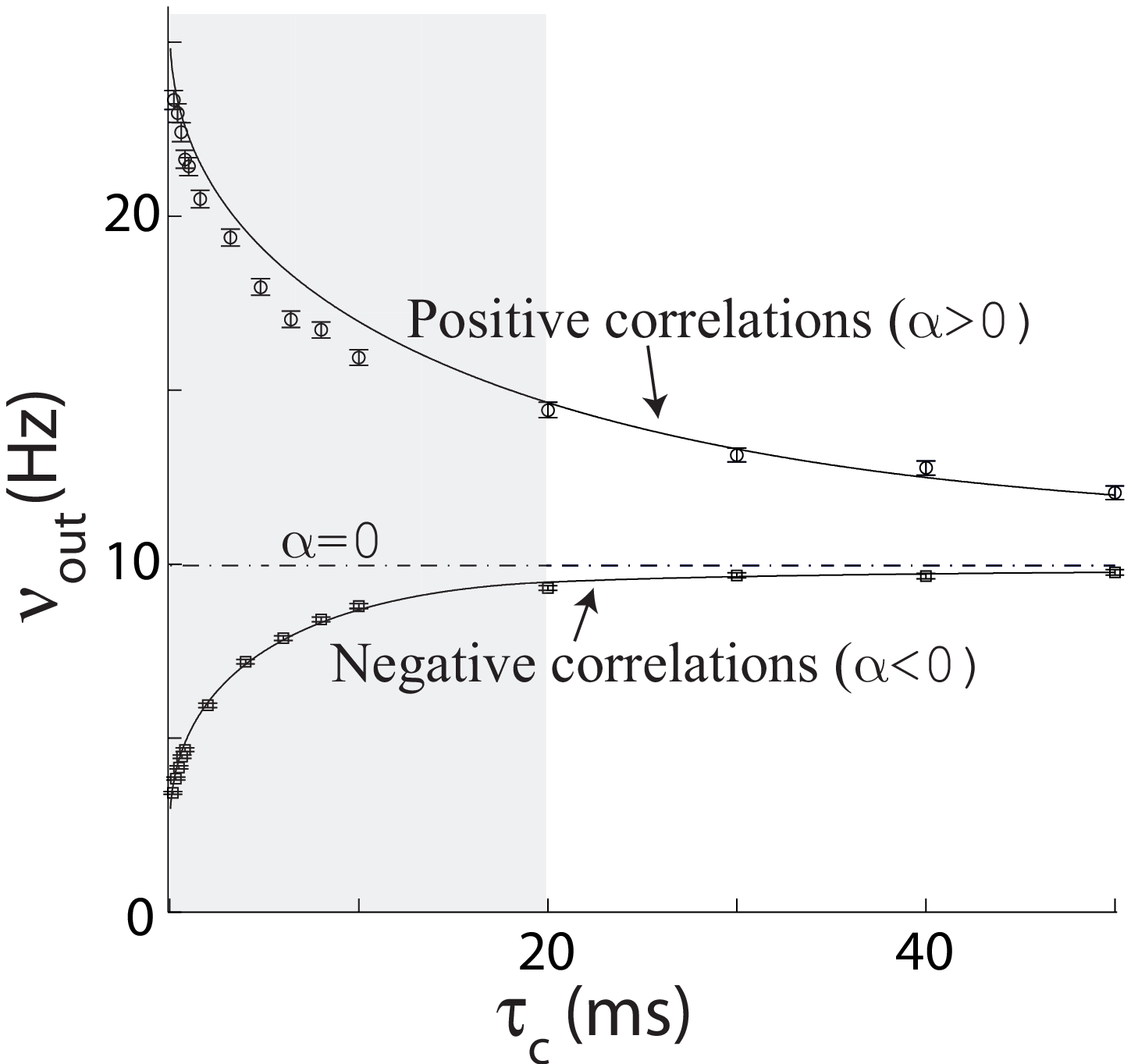}
\end{center}
\caption{\label{fig:grafica1}
}
\end{figure}

\begin{figure}
\begin{center}
\includegraphics[width=12cm,height=7cm,angle=0]{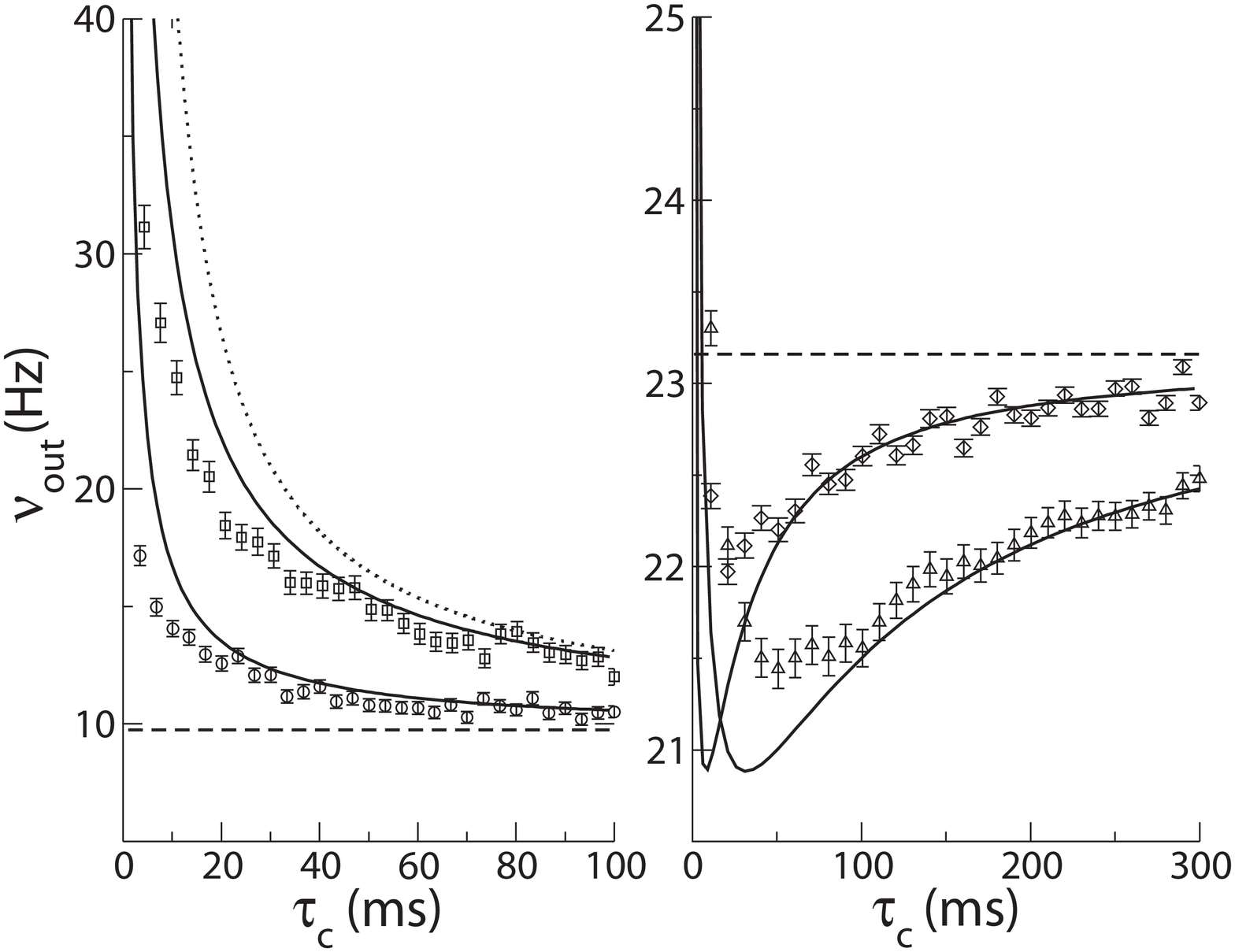}
\end{center}
\caption{\label{fig:grafica-Bal-NoBal}
}
\end{figure}

\begin{figure}
\begin{center}
\includegraphics[width=12cm,height=9cm,angle=0]{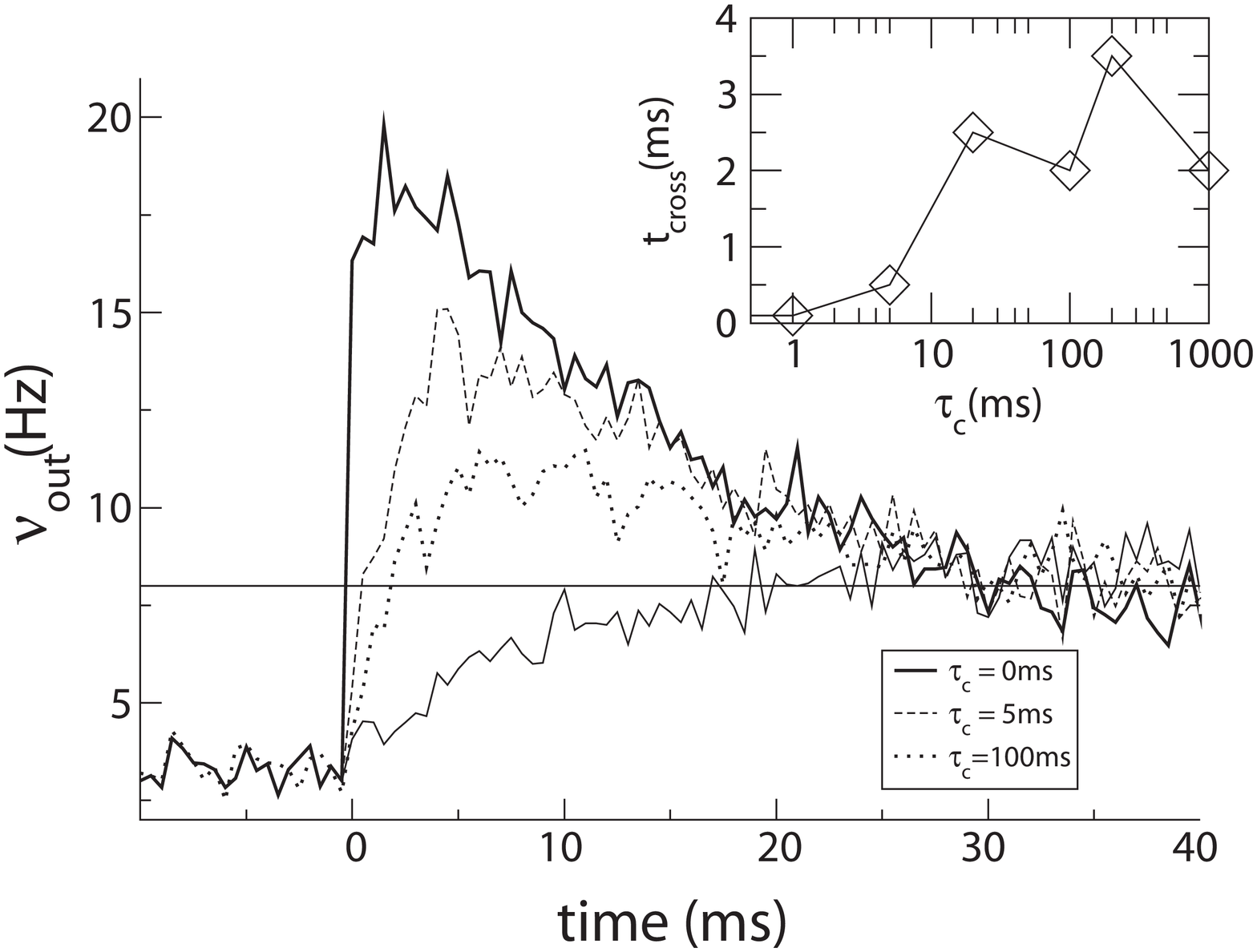}
\end{center}
\caption{\label{fig:histograma1} 
}
\label{fig:histograma1}
\end{figure}


\newpage
Caption 1:
Illustration of the problem studied in this paper 
    (a fully detailed description is given in the text). 
	A set of afferent presynaptic spike trains
	impinges on a LIF neuron. Each individual spike train has
	exponentially-shaped auto-correlations, describing the joint probability density of
	having two spikes separated by a particular time lag (a delta function should 
	be included at zero time lag because the train is made of point events; see text). 
    A fraction of
	the trains also have exponential cross-correlations, describing non-independent firing
	of some of the presynaptic neurons. The total current generated by
    the presynaptic bombardment is replaced by a
	Gaussian process with the same mean and 
    two-point correlation function than that generated
	by the superposition of all presynaptic spike trains. The goal is to characterize
	the spiking response properties of the LIF neuron 
    as a function of the global magnitude and timescale
	of the input correlations. 
	
\newpage
Caption 2:
	(A): An individual afferent spike train from population $p$ 
    could show correlations between two
	times, t and t': the probability of 
    finding a spike at one of those times depends on the
    existence of a spike at the other time. 
	(B): This temporal correlation is described 
    by the auto-correlation function, $C_p(t-t')$,
	assumed to have an exponential shape. The firing rate, $\nu_p$, Fano
	factor, $F_p$, and correlation time, $\tau_c$, 
    enters in the definition of the shape and size of the
	exponential as described in the plot. The delta function present at zero time
	is proportional to $\nu_p$, and participates in the total area of the autocorrelogram.
	(C): When the spike count of the spike train is integrated over a time
	window $T$, the variance of the count divided by $T$ goes exponentially 
    from $\nu_p$ to $F_{N,p}\nu_p$. For small time windows, the count variance 
    converges to that of a Poisson spike train, which is equal to $\nu_p T$. 
    However, for longer time
    windows than the correlation time $\tau_c$, 
	the count variance scales as $F_{N,p}\nu_p$,
    indicating that then the effect of temporal correlations is fully visible.

\newpage
Caption 3:
(A): The probability of having a spike at time $t$ in an afferent spike train 
    belonging to population $p$ could depend on the existence of having a spike at time
    $t'$ on other spike train from population $q$. 
	(B). This correlation is described by the cross-correlation function, $C_{pq}(t-t')$,
	assumed to have exponential shape. The firing rates, Fano
	factors, correlation coefficient of the spike counts, $\rho_{pq}$, 
    and correlation time, $\tau_c$, determine the shape of the exponential,  
	as illustrated in the figure. 
	(C): When the spike counts of the spike trains in the top panel are 
    integrated over a time
	window $T$, their covariance divided by $T$ increases exponentially 
    from zero to a finite value proportional to the correlation coefficient (here we
    define $N'(T)=N(T)-\left<  N(T) \right>$). 
    For short time windows, the covariance is zero 
    and therefore it resembles that
    of two independent spike trains. However, for time windows longer than $\tau_c$,
    correlations are fully visible and the covariance is non-zero.

\newpage
Caption 4:
Diagram of correlations in excitatory ($E$) and inhibitory ($I$) neuronal
  populations presynaptic to the same target neuron. The presynaptic $E$ and $I$ populations 
  make $N_E$ and $N_I$ contacts
  respectively with the target neuron. A fraction $f_{EE(II)}$ of
  these $N_{E(I)}$ excitatory (inhibitory) neurons are correlated with
  each other with a correlation coefficient $\rho_{EE(II)}$.  Also
  there are $E-I$ correlations, with a fraction $f_{EI}$ participating
  from the $E$ population and a fraction $f_{IE}$ from the $I$
  population, for which the correlation
  coefficient is $\rho_{EI} \; (=\rho_{IE})$. Since all $E$ neurons 
  in the fraction $f_{EI}$ are correlated with any given $I$ neuron 
  in the fraction $f_{IE}$, these $E$ neurons necessarily
  have $E-E$ correlations. Therefore, they are considered 
  here to be a group within the fraction
  $f_{EE}$, as shown in the figure. The same applies for the $I$ neurons.

\newpage
Caption 5:
Normalized correlation functions of the current $I(t)$ numerically
  generated by simulating the process defined in eqs.
  (\ref{current} - \ref{dynamic_z}). The normalized correlation function of the current
  is defined as $\hat C_{current}(s)=C_{current}(s)/\sigma_w^2-\delta(s)$,
  where	$C_{current}(s)$ is defined in eq. (\ref{two-point-corr}). The variable
  $s$ is the time lag $s=t-t'$.  With this normalization, the correlation
  function has units of $Hz$. For positive correlations (left) we took $\beta=2$, 
  which yields a correlation
  magnitude $\alpha=8$; $\tau_c = 15 ms$.  For negative correlations (right)
  we took $\beta=-0.5$, which corresponds to $\alpha=-0.75$; 
  here $\tau_c = 5 ms$. In both
  cases, numerical results are compared with the exponential functions
  predicted by eq. (\ref{two-point-corr}) (non-fluctuating curves).

\newpage
Caption 6:
Theoretical predictions (lines) and simulation results (points) 
for the output firing rate of a LIF neuron driven by exponentially correlated inputs
as a function of the correlation timescale.
Here we use eq. (\ref{eq:nu_short_inter}) for short $\tau_c$ and
eq. (\ref{eq:nu_long_inter}) for long $\tau_c$, along with a
continuous and smooth interpolation between the two limits 
(the interpolation is made at an intermediate $\tau_{c,inter} \sim \tau_m$). 
The rate decreases when the 
input correlations are positive ($\alpha>0$, upper curve) and increases when correlations
are negative ($\alpha<0$, lower curve). When there are no correlations ($\alpha=0$),
the neuron fires at a rate of $10Hz$ (dashed-dotted line). Maximum rate differences
relative to the rate with no input correlations are attained when $\tau_c=0$, that is,
when the input correlation is exquisitely precise. Differences are substantial
whenever the correlation time is shorter than the membrane time constant
of the neuron ($\tau_m=20ms$ for this case; shaded region). When the correlation
time becomes longer than $\tau_m$, relative changes are much smaller, and the neuron
becomes less sensitive to the input correlations. Correlation magnitudes are 
$\alpha=8$ (upper curve) and $\alpha=-0.75$ (lower curve), 
and interpolations between the short and long $\tau_c$ theoretical 
predictions were performed at the interpolating time $\tau_{c,inter}=40ms$ 
and $20ms$ respectively. Other parameters are $\tau_{ref}=0ms$, $\Theta=1$ (in
arbitrary units), $H=0$, $\mu=42s^{-1}$, $\sigma^2_w=2s^{-1}$.
Although the short $\tau_c$ expansion requires $\tau_{ref} \neq 0$
the simulation shows that this prediction is good even for zero
$\tau_{ref}$.

\newpage
Caption 7:
Theoretical predictions and simulation results for the firing rate of a LIF
  neuron as function of the correlation timescale 
  for the sub- (left) and the suprathreshold
  regimes (right). Here we use eqs. (\ref{nu_out_big_tauc},\ref{nu_out_big_tauc-gamma}),
  valid for long $\tau_c$. For the subthreshold regime, the
  effect of increasing the correlation time is always to decrease the
  rate. However, for the suprathreshold regime and when the
  input noise is small, the effect is the opposite for long $\tau_c$.
  As the input noise increases, this effect disappears and the
  curve becomes as in the subthreshold regime (data not shown).
  The theoretical predictions (full lines) are
  obtained using the firing rate given in eq.
  (\ref{nu_out_big_tauc-gamma}) without any interpolation, and the
  discrete points are the simulation results with the same parameters
  as in the theoretical curves.  Parameters for the subthreshold
  regime are: $\mu=0 Hz$, $\sigma^2_w=50.5 Hz$, and $\alpha=4$ (top
  full line and squares), $\alpha=1$ (bottom full line and circles) and
  $\alpha=0$ (straight line).  The dotted line has the same parameters
  as the top full line, but it has been obtained from the expression
  of the rate in eq. (\ref{nu_out_big_tauc}). Notice that the prediction
  from eq. (\ref{nu_out_big_tauc-gamma}), strictly only valid for long $\tau_c$,
  is also good even when $\tau_c \sim \tau_m$, and it is better than that
  provided by eq. (\ref{nu_out_big_tauc}) for all $\tau_c$. Parameters for the
  suprathreshold regime are: $\mu=100.7 Hz$, $\sigma^2_w=0.05 Hz$, and
  a very large correlation strength $\alpha=36$ (bottom line and triangles), a moderate
  correlation strength $\alpha=9$ (intermediate line and diamonds) and
  $\alpha=0$ (straight line).  The other parameters are as in Fig.
  (\ref{fig:grafica1}), except for $\tau_m=10ms$.

\newpage
Caption 8:
Averaged transient firing responses of a LIF neuron to changes in the
  input statistics. Below $t=0$ the input is white noise
  ($\alpha=0$) with $\mu=16s^{-1}$ and $\sigma_w^2=0.81s^{-1}$.  Upper
  curve: instantaneous response when $\sigma_w^2$ is increased up to
  $\sigma_w^2=3.8s^{-1}$. Second (third) curve: quick response to
  correlation changes, with $\tau_c=5ms$ ($100ms$) and $\alpha=6.8$
  ($52.3$). Bottom curve: slow response when $\mu$ is changed from
  $\mu=16s^{-1}$to $\mu=19.9s^{-1}$ and $\sigma_w^2$ is kept constant.
  These values were chosen so that the evoked firing rates in the
  final steady state are roughly the same ($\sim 8Hz$, straight line).
  Inset: time when the firing rate response reaches for the first time the
  value of the final stationary rate 
  as a function of $\tau_c$. When the correlation timescale is
  very short, $t_{cross}$ is very small, and it saturates for long
  $\tau_c$.  Neuron parameters are $\tau_m=50ms$, $\tau_{ref}=0$
  ,$\Theta=1$ and $H=0$ (dimensionless).

\newpage


\begin{thebibliography}{65}
\providecommand{\natexlab}[1]{#1}
\providecommand{\url}[1]{\texttt{#1}}
\expandafter\ifx\csname urlstyle\endcsname\relax
  \providecommand{\doi}[1]{doi: #1}\else
  \providecommand{\doi}{doi: \begingroup \urlstyle{rm}\Url}\fi

\bibitem[Abeles(1982)]{Abe+82}
M.~Abeles.
\newblock Role of the cortical neuron: integrator or coincidence detector.
\newblock \emph{Isr. J. Med. Sci.}, 18:\penalty0 83--92, 1982.

\bibitem[Abeles(1991)]{Abe91}
M.~Abeles.
\newblock \emph{Corticonics. Neural circuits of the cerebral cortex.}
\newblock Cambridge UP, Cambridge UK, 1991.

\bibitem[Aersten et~al.(1989)Aersten, Gerstein, Habib, and Palm]{Aer+89}
A.~Aersten, M.~Gerstein, G.~Habib, and G.~Palm.
\newblock Dynamics of neuronal firing correlation: modulation of "effective
  connectivity".
\newblock \emph{J. Neurophysiol.}, 61:\penalty0 900--917, 1989.

\bibitem[Albright(1993)]{Alb93}
T.~D. Albright.
\newblock \emph{Cortical processing of visual motion. In: Visual motion and its
  role in the stabilization of gaze}.
\newblock (Miles F. A., Wallman J., eds.), pag 177-201, New York: Elsevier.,
  1993.
  
\bibitem[Amarasingham et~al.(2006)]{Ama+06}
A. Amarasingham, T. Chen, S. Geman, M.~T. Harrison, D.~L. Sheinberg
\newblock Spike Count Reliability and the Poisson Hypothesis.
\newblock \emph{J. of Neurosci,}, 26(3):\penalty0 801--809, 2006.

\bibitem[Amit and Brunel(1997{\natexlab{a}})]{Ami+97}
D.~J. Amit and N.~Brunel.
\newblock Dynamics of a recurrent network of spiking neurons before and
  following learning.
\newblock \emph{Network}, 8:\penalty0 373, 1997{\natexlab{a}}.

\bibitem[Amit and Brunel(1997{\natexlab{b}})]{Ami+97-spont}
D.~J. Amit and N.~Brunel.
\newblock Model of global spontaneous activity and local structured activity
  during delay periods in the cerebral cortex.
\newblock \emph{Cereb. Cortex}, 7:\penalty0 237--52, 1997{\natexlab{b}}.

\bibitem[Averbeck and Lee(2004)]{Averbeck+04-syn}
B.~B. Averbeck and D.~Lee.
\newblock Coding and transmission of information by neural ensembles.
\newblock \emph{Trends in Neuroscience}, 27(4):\penalty0 225--30, 2004.

\bibitem[Bair et~al.(2001)Bair, Zohary, and Newsome]{Bair+01-tauc}
W.~Bair, E.~Zohary, and W.~T. Newsome.
\newblock Correlated firing in macaque visual area mt: Time scales and
  relationship to behavior.
\newblock \emph{J. of Neurosci,}, 21(5):\penalty0 1676--97, 2001.

\bibitem[Bernander et~al.(1991)Bernander, Douglas, Martin, and Koch]{Ber+91}
O.~Bernander, R.~J. Douglas, K.~A. Martin, and C.~Koch.
\newblock Synaptic background activity influences spatiotemporal integration in
  single pyramidal cells.
\newblock \emph{Proc. Natl. Acad. Sci. USA}, 88:\penalty0 11569--11573, 1991.

\bibitem[Braitenberg and Sch{\"u}z(1991)]{Bra+91}
V.~Braitenberg and A.~Sch{\"u}z.
\newblock \emph{Anatomy of the Cortex: Statistics and Geometry}.
\newblock Springer Verlag, Berlin, 1991.

\bibitem[Brunel and Sergi(1998)]{Bru+98b}
N.~Brunel and S.~Sergi.
\newblock Firing frequency of leaky integrate-and-fire neurons with synaptic
  current dynamics.
\newblock \emph{J. Theor. Biol.}, 195:\penalty0 87--95, 1998.

\bibitem[Burkitt and Clark(1999)]{Bur+99}
A.~N. Burkitt and G.~M. Clark.
\newblock Analysis of integrate-and-fire neurons: synchronization of synaptic
  input and spike output.
\newblock \emph{Neural Comput.}, 11:\penalty0 871--901, 1999.

\bibitem[Cateau and Reyes(2006)]{Cateau+06-propag}
H.~Cateau and A.~Reyes.
\newblock Relation between single neuron and population spiking statistics and
  effects on network activity.
\newblock \emph{Phys. Rev. Lett.}, 96:\penalty0 058101, 2006.

\bibitem[Compte et~al.(2003)Compte, Constantinidis, Tegner, Raghavachari,
  Chafee, Goldman-Rakic, and Wang]{Compte+03-F_N}
A.~Compte, C.~Constantinidis, J.~Tegner, S.~Raghavachari, M.~V. Chafee, P.~S.
  Goldman-Rakic, and X.-J. Wang.
\newblock Temporally irregular mnemonic persistent activity in prefrontal
  neurons of monkeys during a delayed response task.
\newblock \emph{J Neurophysiol}, 90:\penalty0 3441--3454, 2003.

\bibitem[Cragg(1967)]{Cra67}
B.~G. Cragg.
\newblock The density of synapses and neurones in the motor and visual areas of
  the cerebral areas.
\newblock \emph{J. Anat.}, 101:\penalty0 639--654, 1967.

\bibitem[Daley and Vere-Jones(1988)]{Daley+88}
D.~J. Daley and D.~Vere-Jones.
\newblock \emph{An introduction to the theory of point processes}.
\newblock Springer, New York, 1988.

\bibitem[Dean(1981)]{Dea81}
A.~F. Dean.
\newblock The variability of discharge of simple cells in cat striate cortex.
\newblock \emph{Exp. Brain Res.}, 44:\penalty0 437--440, 1981.

\bibitem[deCharms and Merzenich(1996)]{deC+96}
R.~C. deCharms and M.~M. Merzenich.
\newblock Primary cortical representation of sounds by the coordination of
  action potentials.
\newblock \emph{Nature}, 381:\penalty0 610--613, 1996.

\bibitem[DeFelipe and Fari{\~n}as(1992)]{DeF+92}
J.~DeFelipe and I.~Fari{\~n}as.
\newblock The pyramidal neuron of the cerebral cortex: morphological and
  chemical characteristics of the synaptic inputs.
\newblock \emph{Prog. Neurobiol.}, 39:\penalty0 563--607, 1992.

\bibitem[Doering et~al.(1987)Doering, Hagan, and Levermore]{Doe+87}
C.~R. Doering, P.~S. Hagan, and C.~D. Levermore.
\newblock Bistability driven by weakly colored gaussian noise: the
  fokker-planck equation boundary layer and mean first-passage times.
\newblock \emph{Physical Review Letters}, 59 (19):\penalty0 2129--2132, 1987.

\bibitem[Doiron et~al.(2006)Doiron, Rinzel, and Reyes]{Doiron+06}
B.~Doiron, J.~Rinzel, and A.~Reyes.
\newblock Stochastic synchronization in finite size spiking networks.
\newblock \emph{Physical Review E}, 74:\penalty0 030903, 2006.

\bibitem[Feng and Brown(2000)]{Fen+00}
J.~Feng and D.~Brown.
\newblock Impact of correlated inputs on the output of the integrate-and-fire
  model.
\newblock \emph{Neural Computation}, 12:\penalty0 671--692, 2000.

\bibitem[Fries et~al.(1997)Fries, Roelfsema, Engel, Konig, and
  Singer]{Fries+97}
P.~Fries, P.~R. Roelfsema, A.~K. Engel, P.~Konig, and W.~Singer.
\newblock Synchronization of oscillatory responses in visual cortex correlates
  with perception in interocular rivalry.
\newblock \emph{Proc. Natl. Acad. Sci}, 94:\penalty0 12699--704, 1997.

\bibitem[Fries et~al.(2001)Fries, Reynolds, Rorie, and Desimone]{Fri+01}
P.~Fries, J.~H. Reynolds, A.~E. Rorie, and R.~Desimone.
\newblock Modulation of oscillatory neuronal synchronization by selective
  visual attention.
\newblock \emph{Science}, 291:\penalty0 1560--1563, 2001.

\bibitem[Gochin et~al.(1991)Gochin, Miller, Gross, and Gerstein]{Goc+91}
P.~M. Gochin, E.~K. Miller, C.~G. Gross, and G.~L. Gerstein.
\newblock Functional interactions among neurons in inferior temporal cortex of
  the awake macaque.
\newblock \emph{Exp. Brain Res.}, 84:\penalty0 505--516, 1991.

\bibitem[Kuhn et~al.(2003)Kuhn, Aertsen, and Rotter]{Kuhn+03}
A.~Kuhn, A.~Aertsen, and S.~Rotter.
\newblock Higher-order statistics of input ensembles and the response of simple
  models neurons.
\newblock \emph{Neural Computation}, 15:\penalty0 67--101, 2003.

\bibitem[LaCamera et~al.(2004)LaCamera, Rauch, Luscher, Senn, and
  Fusi]{LaCamera+04}
G.~LaCamera, A.~Rauch, H.-R. Luscher, W.~Senn, and S.~Fusi.
\newblock Minimal models of adapted neuronal response to in vivo–like input
  currents.
\newblock \emph{Neural Computation}, 16:\penalty0 2101--2124, 2004.

\bibitem[Laurent(2001)]{Lau+01}
G.~Laurent.
\newblock Odor encoding as an active, dynamical process: experiments,
  computation and theory.
\newblock \emph{Ann. Rev. Neurosci.}, 24:\penalty0 263--297, 2001.

\bibitem[Lee et~al.(1998)Lee, Port, Kruse1, and Georgopoulos]{Lee+98-var}
D.~Lee, N.~L. Port, W.~Kruse1, and A.~P. Georgopoulos.
\newblock Variability and correlated noise in the discharge of neurons in motor
  and parietal areas of the primate cortex.
\newblock \emph{The Journal of Neuroscience}, 18(3):\penalty0 1161--70, 1998.

\bibitem[Lerchner et~al.(2006)Lerchner, Ursta, Hertz, Ahmadi, Ruffiot, and
  Enemark]{Lerchner+06}
A.~Lerchner, C.~Ursta, J.~Hertz, M.~Ahmadi, P.~Ruffiot, and S.~Enemark.
\newblock Response variability in balanced cortical networks.
\newblock \emph{Neural Computation}, 18:\penalty0 634--659, 2006.

\bibitem[Lindner(2006)]{Lindner06}
B.~Lindner.
\newblock Superposition of many independent spike trains is generally not a
  poisson process.
\newblock \emph{Phys. Rev. E}, 73(2):\penalty0 022901, 2006.

\bibitem[Lindner et~al.(2005)Lindner, Doiron, and Longtin]{Lindner+05}
B.~Lindner, B.~Doiron, and A.~Longtin.
\newblock Theory of oscillatory firing induced by spatially correlated noise
  and delayed feedback.
\newblock \emph{Physical Review E}, 72:\penalty0 061919, 2005.

\bibitem[Masuda(2006)]{Masuda06}
N.~Masuda.
\newblock Simultaneous rate-synchrony codes in populations of spiking neurons.
\newblock \emph{Neural Computation}, 18:\penalty0 45--59, 2006.

\bibitem[Moreno and Parga(2002)]{Mor+02a}
R.~Moreno and N.~Parga.
\newblock \emph{Firing rate for a generic integrate-and-fire neuron with
  exponentially correlated input.}
\newblock 223-8, In Lecture notes in computer science. Ed. J.R. Dorronsoro.
  Springer Verlag, 2002.

\bibitem[Moreno et~al.(2002)Moreno, de~la Rocha, Renart, and Parga]{Mor+02}
R.~Moreno, J.~de~la Rocha, A.~Renart, and N.~Parga.
\newblock Response of spiking neurons to correlated inputs.
\newblock \emph{Physical Review Letters}, 89 (28):\penalty0 288101, 2002.

\bibitem[Moreno-Bote and Parga(2004)]{Mor+04}
R.~Moreno-Bote and N.~Parga.
\newblock Role of synaptic filtering on the firing response of simple model
  neurons.
\newblock \emph{Physical Review Letters}, 92(2):\penalty0 028102, 2004.

\bibitem[Moreno-Bote and Parga(2005)]{Mor+05}
R.~Moreno-Bote and N.~Parga.
\newblock Membrame potential and response properties of populations of cortical
  neurons in the high conductance state.
\newblock \emph{Physical Review Letters}, 94:\penalty0 088103, 2005.

\bibitem[Moreno-Bote and Parga(2006)]{Mor+06}
R.~Moreno-Bote and N.~Parga.
\newblock Auto- and crosscorrelograms for the spike response of leaky
  integrate-and-fire neurons with slow synapses.
\newblock \emph{Physical Review Letters}, 96:\penalty0 028101, 2006.

\bibitem[Nowak et~al.(1999)Nowak, Munk, James, Girard, and
  Bullier]{Nowak+99-corr}
L.~G. Nowak, M.~H.~J. Munk, A.~C. James, P.~Girard, and J.~Bullier.
\newblock Cross-correlation study of the temporal interactions between areas v1
  and v2 of the macaque monkey.
\newblock \emph{J. Neurophysiol.}, 81:\penalty0 1057--74, 1999.

\bibitem[Nykamp and Tranchina(2001)]{Nykamp+01}
D.~Nykamp and D.~Tranchina.
\newblock A population density approach that facilitates large-scale modeling
  of neural networks: extension to slow inhibitory synapses.
\newblock \emph{Neural Computation}, 13:\penalty0 511--546, 2001.

\bibitem[Perkel et~al.(1967)Perkel, Gerstein, and Moore]{Per+67}
D.~H. Perkel, G.~L. Gerstein, and G.~P. Moore.
\newblock Neuronal spike trains and stochastic point processes. ii.
  simulataneous spike trains.
\newblock \emph{Biophys. J.}, 7:\penalty0 419--440, 1967.

\bibitem[Renart et~al.(2001)Renart, Moreno, de~la Rocha, Rolls, and
  Parga]{Ren+01}
A.~Renart, R.~Moreno, J.~de~la Rocha, E.~Rolls, and N.~Parga.
\newblock A model of the it-pf network in object working memory which includes
  balanced persistent activity and tuned inhibition.
\newblock \emph{Neurocomputing}, 28:\penalty0 1525--1531, 2001.

\bibitem[Renart et~al.(2003)Renart, Brunel, and Wang]{Renart+03book}
A.~Renart, N.~Brunel, and X.~J. Wang.
\newblock \emph{Mean-field theory of recurrent cortical networks: From
  irregularly spiking neurons to working memory. In J. Feng (Ed.),
  Computational neuroscience: A comprehensive approach}.
\newblock CRC Press, Boca Raton, FL, 2003.

\bibitem[Renart et~al.(2007)Renart, Moreno-Bote, Wang, and Parga]{Renart+07}
A.~Renart, R.~Moreno-Bote, X.-J. Wang, and N.~Parga.
\newblock Mean-driven and fluctuation-driven persistent activity in recurrent
  networks.
\newblock \emph{Neural Computation}, 19:\penalty0 1--46, 2007.

\bibitem[Ricciardi(1977)]{Ric77}
L.~M. Ricciardi.
\newblock \emph{Diffusion processes and related topics in biology}.
\newblock Springer-Verlag, Berlin, 1977.

\bibitem[Richardson and Gerstner(2005)]{Richardson+05}
M.~Richardson and W.~Gerstner.
\newblock Synaptic shot noise and conductance fluctuations affect the membrane
  voltage with equal significance.
\newblock \emph{Neural Computation}, 17:\penalty0 923--947, 2005.

\bibitem[Riehle et~al.(1997)Riehle, Grun, Diesmann, and Aertsen]{Rie+97}
A.~Riehle, S.~Grun, M.~Diesmann, and A.~Aertsen.
\newblock Spike synchronization and rate modulation differentially involved in
  motor cortical function.
\newblock \emph{Science}, 278:\penalty0 1950--1953, 1997.

\bibitem[Risken(1989)]{Ris89}
H.~Risken.
\newblock \emph{The Fokker-Planck equation. 2$^{nd}$ Ed.}
\newblock Springer-Verlag, Berlin, 1989.

\bibitem[Rolls and Treves(1998)]{Rol+98b}
E.~T. Rolls and A.~Treves.
\newblock \emph{Neural networks and brain function}.
\newblock Oxford University Press, Oxford, U.K., 1998.

\bibitem[Rudolph and Destexhe(2001)]{Rud+01}
M.~Rudolph and A.~Destexhe.
\newblock Correlation detection and resonance in neural systems with
  distributed noise sources.
\newblock \emph{Phyical Review Letters}, 86(16):\penalty0 3662--4, 2001.

\bibitem[Salinas and Sejnowski(2000)]{Sal+00}
E.~Salinas and T.~J. Sejnowski.
\newblock Impact of correlated synaptic input on output firing rate and
  variability in simple neuronal models.
\newblock \emph{J. Neurosci.}, 20:\penalty0 6193--6209, 2000.

\bibitem[Salinas and Sejnowski(2001)]{Sal+01a}
E.~Salinas and T.~J. Sejnowski.
\newblock Correlated neuronal activity and the flow of neural information.
\newblock \emph{Nature Reviews Neuroscience}, 2:\penalty0 539--550, 2001.

\bibitem[Shadlen and Newsome(1998)]{Sha+98}
M.~N. Shadlen and W.~T. Newsome.
\newblock The variable discharge of cortical neurons: implications for
  connectivity, computation, and information coding.
\newblock \emph{J. Neurosci.}, 18:\penalty0 3870--3896, 1998.

\bibitem[Silberberg et~al.(2004)Silberberg, Bethge, Markram, Pawelzik, and
  Tsodyks]{Silberberg+04}
G.~Silberberg, M.~Bethge, H.~Markram, K.~Pawelzik, and M.~Tsodyks.
\newblock Dynamics of population rate codes in ensembles of neocortical
  neurons.
\newblock \emph{Journal of Neurophysiology}, 91:\penalty0 704--709, 2004.

\bibitem[Softky(1994)]{Sof94}
W.~Softky.
\newblock Submillisecond coincidence detection in active dendritic trees.
\newblock \emph{Neuroscience}, 58:\penalty0 13--41, 1994.

\bibitem[Softky and Koch(1993)]{Sof+93}
W.~Softky and C.~Koch.
\newblock The highly irregular firing of cortical cells is incosistent with
  temporal integration of random epsp's.
\newblock \emph{J. Neurosci.}, 13:\penalty0 334--350, 1993.

\bibitem[Steinmetz et~al.(2000)Steinmetz, Roy, Fitzgerald, Hsiao, Johnson, and
  Niebur]{Ste+00}
P.~N. Steinmetz, A.~Roy, P.~J. Fitzgerald, S.~S. Hsiao, K.~O. Johnson, and
  E.~Niebur.
\newblock Attention modulates synchronized neuronal firing in primate
  somatosensory cortex.
\newblock \emph{Nature}, 404:\penalty0 187--190, 2000.

\bibitem[Stevens and Zador(1998)]{Ste+98}
C.~F. Stevens and A.~M. Zador.
\newblock Input synchrony and the irregular firing of cortical neurons.
\newblock \emph{Nature Neurosci.}, 1(3):\penalty0 210--217, 1998.

\bibitem[Ts'o et~al.(1986)Ts'o, Gilbert, and Wiesel]{Tso+86}
D.~Y. Ts'o, C.~D. Gilbert, and T.~N. Wiesel.
\newblock Relationships between horizontal interactions and functional
  architecture in cat striate cortex as reveales by cross-correlations
  analysis.
\newblock \emph{J. Neuroscie.}, 6:\penalty0 1160--1170, 1986.

\bibitem[Tuckwell(1988)]{Tuc88}
H.~C. Tuckwell.
\newblock \emph{Introduction to theoretical neuroscience. Vol. 1 and 2}.
\newblock Cambridge UP, Cambridge UK, 1988.

\bibitem[Usrey and Reid(1999)]{Unrey+99-syn}
W.~M. Usrey and R.~C. Reid.
\newblock Synchronous activity in the visual system.
\newblock \emph{Annu. Rev. Physiol.}, 61:\penalty0 435--56, 1999.

\bibitem[Vaadia et~al.(1995)Vaadia, Haalman, Abeles, Bergman, Prut, Slovin, and
  Aertsen]{Vaadia+95}
E.~Vaadia, I.~Haalman, M.~Abeles, H.~Bergman, Y.~Prut, H.~Slovin, and
  A.~Aertsen.
\newblock Dynamics of neuronal interactions in monkey cortex in relation to
  behavioural events.
\newblock \emph{Nature}, 373:\penalty0 515--518, 1995.

\bibitem[Wehr and Laurent(1999)]{Weh+99}
M.~Wehr and G.~Laurent.
\newblock Relationship between afferent and central temporal patterns in the
  locust olfatory system.
\newblock \emph{J. Neurosci.}, 19:\penalty0 381--390, 1999.

\bibitem[White(1989)]{Whi89}
E.~L. White.
\newblock Cortical circuits.
\newblock \emph{Birkauser}, 1989.

\bibitem[Zohary et~al.(1994)Zohary, Shadlen, and Newsome]{Zoh+94}
E.~Zohary, M.~N. Shadlen, and W.~T. Newsome.
\newblock Correlated neuronal discharge rate and its implication for
  psychophysical performance.
\newblock \emph{Nature}, 370:\penalty0 140--143, 1994.

\end{thebibliography}

\end{document}